\documentclass[fleqn,usenatbib]{mnras}

\usepackage{newtxtext,newtxmath}

\usepackage[T1]{fontenc}
\usepackage{booktabs}
\DeclareRobustCommand{\VAN}[3]{#2}
\let\VANthebibliography\thebibliography
\def\thebibliography{\DeclareRobustCommand{\VAN}[3]{##3}\VANthebibliography}

\usepackage{graphicx}	
\usepackage{amsmath}	
\usepackage [latin1]{inputenc}

\title[Cometary ion dynamics at 67P]{Cometary ion dynamics at 67P: A collisional test-particle approach with Rosetta data comparison}

\author[Z. M. Lewis et al.]{
Z. M. Lewis,$^{1,2}$\thanks{E-mail: zoe.lewis@lancaster.ac.uk}
P. Stephenson,$^{3}$
A. Beth,$^{1}$
M. Galand,$^{1}$
E. Kallio,$^{4}$
and A. Moeslinger$^5$
\\
$^{1}$Department of Physics, Imperial College London, UK\\
$^{2}$Lancaster University, Bailrigg, UK\\
$^{3}$Lunar and Planetary Laboratory, Tucson, AZ, USA\\
$^{4}$Aalto University, Helsinki, Finland \\
$^{5}$Ume\aa~University, Ume\aa, Sweden
}

\date{Accepted XXX. Received YYY; in original form ZZZ}

\pubyear{\the\year{}}

\begin{document}
\label{firstpage}
\pagerange{\pageref{firstpage}--\pageref{lastpage}}
\maketitle

\begin{abstract}
The Rosetta spacecraft escorted comet 67P/Churyumov-Gerasimenko for two years, gathering a rich and variable dataset. Amongst the data from the Rosetta Plasma Consortium (RPC) suite of instruments are measurements of the total electron density from the Mutual Impedance Probe (MIP) and Langmuir Probe (LAP). 

At low outgassing, the plasma density measurements can be explained by a simple balance between the production through ionisation and loss through transport. Ions are assumed to travel radially at the outflow speed of the neutral gas. Near perihelion, the assumptions of this field-free chemistry-free model are no longer valid, and plasma density is overestimated. This can be explained by enhanced ion transport by an ambipolar electric field inside the diamagnetic cavity, where the interplanetary magnetic field does not reach.

In this study, we explore the transition between these two regimes, at intermediate outgassing ($5.4 \times10^{26}~\mathrm{s^{-1}}$), when the interaction between the cometary and solar wind plasma influences the transport of the ions. We use a 3D collisional test-particle model, adapted from \citet{Stephenson2022} to model the cometary ions with input electric and magnetic fields from a hybrid simulation for 2.5-3 au. The total plasma density from this model is then compared to data from MIP/LAP and to the field-free chemistry-free model. In doing so, we highlight the limitations of the hybrid approach and demonstrate the importance of modelling collisional cooling of the electrons to understand the ion dynamics close to the nucleus.

\end{abstract}




\section{Introduction}

The cometary ionosphere is formed by the ionisation of the neutral coma, expanding neutral gas released by sublimating ices from the nucleus. Unlike at planets, these ions are not gravitationally bound, but are instead continuously produced and lost to space. 

The ion environment of Comet 67P/Churyumov-Gerasimenko was explored extensively by the ESA Rosetta mission during its main escort phase \citep{Glassmeier2007b}. Amongst the large and varied payload was the Rosetta Plasma Consortium (RPC) suite of instruments \citep{Carr2007}, designed to measure the charged particle populations and to shed light on the comet-solar wind interaction. In this study, we use electron density data from the Mutual Impedance Probe (MIP, \cite{Trotignon2007}) and the Langmuir Probe (LAP, \cite{Eriksson2007}).

\subsection{Ionospheric chemistry}

 For most of the Rosetta escort, the neutral coma of 67P was primarily comprised of water, that can be directly ionised by solar photons, electron impact, or solar wind charge exchange, to produce $\mathrm{H_2O^+}$. When the coma is dense enough, $\mathrm{H_2O}^+$ is quickly lost through the transfer of a proton to $\mathrm{H_2O}$, forming $\mathrm{H_3O}^+$:
\begin{equation}
    \mathrm{H_2O} + \mathrm{H_2O^+} \to \mathrm{H_3O^+} + \mathrm{OH}.
    \label{eq: PT water}
\end{equation}
This reaction happens so readily that $\mathrm{H_3O^+}$ is often the dominant ion species in the cometary coma (e.g. \cite{Murad1987}, \cite{Heritier2017a}). If the outgassing is high enough, $\mathrm{H_3O}^+$ can then be lost to reactions with high proton affinity neutrals (\cite{Altwegg1993}, \cite{Vigren2013}, \cite{Heritier2017a}). These are neutrals for which it is energetically favourable to `steal' a proton from $\mathrm{H_3O}^+$, and this process can happen repeatedly for neutrals with increasing proton affinity until the terminal ion, whose associated neutral has the highest proton affinity, is reached. The terminal ion is then lost through transport or ion-electron dissociative recombination. The neutral with the highest proton affinity in the coma of 67P is $\mathrm{NH_3}$, leading to the formation of its protonated version, $\mathrm{NH_4^+}$ (\cite{Altwegg1993}, \cite{Vigren2013}):
\begin{equation}
    \mathrm{NH_3} + \mathrm{H_3O^+} \to \mathrm{NH_4^+} + \mathrm{H_2O}.
\end{equation}

$\mathrm{NH_4^+}$ can also be produced through interaction between $\mathrm{NH_3}$ and other protonated versions of neutrals with proton affinity higher than water but lower than $\mathrm{NH_3}$. Therefore, although $\mathrm{H_3O^+}$ is usually the dominant ion species, when the coma is dense enough for ion-neutral chemistry to happen more readily, $\mathrm{NH_4^+}$ can overtake it. Only a small mixing ratio of $\mathrm{NH_3}$ in the neutral population is required for significant production of $\mathrm{NH_4^+}$. Since it can only be produced through ion-neutral chemistry, and not directly through ionisation, $\mathrm{NH_4^+}$ is a useful indicator that the timescale for ion-neutral chemistry is similar to or less than the timescale for ion transport (\cite{Vigren2013}, \cite{Heritier2017a}, \cite{Beth2019}). $\mathrm{NH_4^+}$ can hence be used as a tracer of ion-neutral chemistry occurring in the coma 

$\mathrm{NH_4^+}$ has been detected at 67P by the ROSINA-DFMS (Rosetta Orbiter Spectrometer for Ion and Neutral Analysis - Double-Focussing Mass Spectrometer, \cite{Balsiger2007}) using the high mass-resolution mode \citep{Beth2016}. Detections increase with outgassing, and the ion is more prevalent near perihelion (\cite{Heritier2017a}, \cite{Lewis2023}), where the coma is more dense and ion-neutral reactions can happen more efficiently. \citet{Lewis2023} showed that DFMS counts of $\mathrm{NH_4^+}$ are greater inside the diamagnetic cavity - the magnetic field-free region surrounding the nucleus, sporadically detected by the RPC magnetometer (MAG, \cite{Glassmeier2007}, \cite{Goetz2016a}). This correlation between the diamagnetic cavity and $\mathrm{NH_4^+}$ demonstrates the link between the plasma dynamics and the ion composition. 

\subsection{Plasma density driven by ion dynamics}

The simplest parameters we can use to describe the cometary ion population are the total ion density $n_i$ (which is approximately equal to the total electron density $n_e$, since quasi-neutrality is maintained in the plasma), and the ion bulk velocity $\vec{u_i}$. These parameters are inextricably linked, and tied together by the continuity equation
\begin{equation}
    \frac{\partial{n_{i}}}{\partial{t}} + \mathbf{\nabla} \cdot (n_{i}\vec{u}_{i}) = (\nu^{h\nu} + \nu^{e^-})n_n - R_{DR}n_e
    \label{eq:continuity}
\end{equation}
where $\nu^{h\nu}$ and $\nu^{e^-}$ are the photo-ionisation and electron-impact ionisation frequencies, respectively. $n_n$ is the density of the neutral coma, which can be expressed by a simplified Haser model \citep{Haser1957} as $n_n = Q/(4\pi u_n r^2)$, where $u_n$ is the neutral expansion velocity and $Q$, $m$ the total outgassing rate. $R_{DR}$ is the effective loss rate of the ions due to ion-electron dissociative recombination. 

\begin{figure*}
    \includegraphics[width=\linewidth]{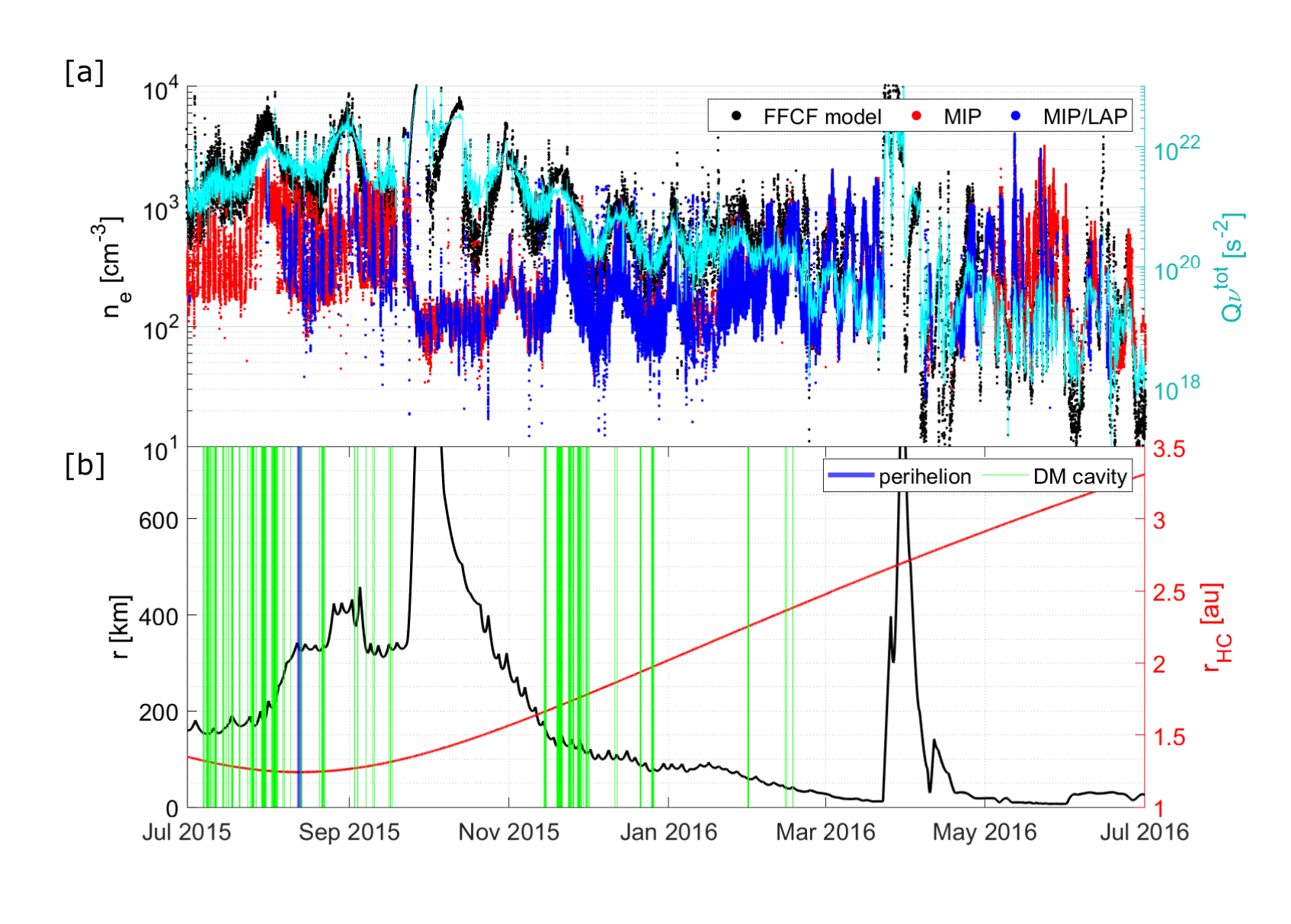}
    \caption{ [a] Comparison of the (black) calculated plasma density from Eq. \ref{eq: FFCF}, with the measured electron density from (red) RPC-MIP and (blue) the MIP/LAP combined dataset \protect\citep{Johansson2021}. (cyan) The ion production rate $Q\nu^{\mathrm{tot}}$ where $Q$ is the total outgassing rate calculated COPS and $\nu^{\mathrm{tot}}$ is the sum of the total photoionisation and electron-impact frequencies \protect\citep{Stephenson2023a}. Data are shown for July 2015 - July 2016. [b] (black) cometocentric distance, and (red) heliocentric distance. Green lines mark diamagnetic cavity crossings (\protect\cite{Goetz2016a}, width of line not to scale) and blue line indicates perihelion in August 2015.}
    \label{fig: plas dens motivation}
\end{figure*}

For large heliocentric distances and close to the comet, \citet{Galand2016} demonstrated that a simple balance between the ionisation rate and radial ion transport was sufficient to explain the RPC--MIP and RPC--LAP electron density. For such a low outgassing (at $<3$~au), the following conditions were determined to be valid:
\begin{itemize}
    \item No attenuation of the incoming solar EUV (coma is optically thin),
    \item No plasma loss through dissociative recombination ($R_{DR}=0$)
\end{itemize}
Further to this, a simplifying assumption was made: that the cometary ions travel radially outwards at the same speed as the neutrals. Under these conditions the continuity equation (Eq. \ref{eq:continuity}) can be solved to find that the ion density $n_i$ is simply
\begin{equation}
    n_i = \frac{(\nu^{h\nu} + \nu^{e^-}) n_n}{u_n} (r-r_c)
    \label{eq: FFCF}
\end{equation}
where $r_c$ is the approximate radius of the nucleus. A comparison of the plasma density calculated using Eq.~\ref{eq: FFCF} and from RPC/MIP and RPC/LAP measurements is shown in Figure~\ref{fig: plas dens motivation}. A similar comparison was made in \citet{Vigren2019}, but we now present an updated version with the combined LAP/MIP dataset \citep{Johansson2021} and including both photo- and electron-impact ionisation frequencies \citep{Stephenson2023a}.

In general, this field-free chemistry-free model works well to explain the electron density at low outgassing post-perihelion (\cite{Heritier2017b}, \cite{Heritier2018a}) as well as pre-perihelion \citep{Galand2016}. The range of cometocentric distances probed by Rosetta at this time was usually within a few tens of kilometres from the surface, so the full parameter space could not be explored, but there is good agreement over the regions probed, which extended up to $\sim$~80~km post-perihelion \citep{Heritier2018a} (see Figure~\ref{fig: plas dens motivation} after March 2016).  

At higher outgassing ($Q\gtrsim 10^{27}~\mathrm{s^{-1}}$), the plasma density is no longer well constrained by the simple model (see \cite{Vigren2019} and Figure~\ref{fig: plas dens motivation}). This is most likely due to the ions being accelerated by the ambipolar electric field such that $u_i \neq u_n$ (e.g. \cite{Vigren2017b}). \citet{Lewis2024} showed through a 1D ionospheric model that acceleration of the ions by an electric field proportional to $1/r$, could be used to bring the modelled densities in line with those observed by RPC--MIP and RPC--LAP within the diamagnetic cavity. Outside the diamagnetic cavity, the non-zero magnetic field contributes to the convective and Hall terms (see e.g. \cite{Gunell2019}, \cite{Beth2017a}, \cite{Deca2019}) into the electric field, and the flow is no longer simply radial. 

As demonstrated in \citet{Vigren2019}, a transition occurs during late February to early March 2016 where the assumptions of the simple model become valid. Interestingly, the transition appears to coincide with the last observations of the diamagnetic cavity by RPC-MAG \citep{Goetz2016a} in mid February 2016. It is not known whether the diamagnetic cavity boundary lay below the spacecraft beyond these dates, but given the relatively short (and decreasing) distance of Rosetta from the comet nucleus it seems unlikely that the cavity persisted into March 2016.

Direct determination of the bulk ion velocity from Rosetta RPC measurements is possible, though challenging due the effect of the strongly negative spacecraft potential \citep{Odelstad2017}, and uncertainty in the ion temperature. Values of 2--8~km\,s$^{-1}$ have been derived using the current-voltage characteristics from RPC--LAP and the plasma density from RPC--MIP, using diamagnetic cavity crossings in August 2015 \citep{Vigren2017b} and November 2015 \citep{Odelstad2018}. \citet{Bergman2021} used RPC--ICA energy spectra from 88 diamagnetic cavity crossings to derive bulk speed of 5--10~km\,s$^{-1}$, however the spacecraft potential distorts the measured direction and velocity of low energy ions from ICA, requiring disentangling with Particle-In-Cell simulations \citep{Bergman2021a}. \citet{Williamson2024b} used a statistical approach, fitting drifting Maxwell-Boltzmann distributions to over 3000 RPC--ICA scans from across the whole Rosetta escort phase. They found that this produced consistently higher ion energies than those derived from RPC--LAP.

Despite some variations, all of these measurements suggest that the ion speed is higher than the  neutral expansion speed $u_n$ (0.5--1~km\,s$^{-1}$, \cite{Biver2019}, \cite{Hansen2016}), suggesting collisional decoupling between the ions and neutrals, and electric field acceleration of the ions. From 1D ion acceleration model simulations, \citet{Lewis2024} found that an electric field in $r^{-1}$ of around $2~\mathrm{mV~m^{-1}}$ at the surface was sufficient to reproduce the plasma densities measured by RPC inside the diamagnetic cavity (for July and November 2015 case studies). Such a field would lead to bulk ion speeds at Rosetta of $1.4-3.0~\mathrm{km~s^{-1}}$ in the diamagnetic cavity. However, the good agreement of the field-free chemistry-free model (Eq.~\ref{eq: FFCF}) at lower outgassing (\cite{Galand2016}, \cite{Heritier2018a}) would suggest that an ion flow much closer to the neutral speed is dominant at larger heliocentric distances. One might expect that electric and magnetic fields are more likely to effect the ion velocity in the low outgassing regime, when there is no diamagnetic cavity and the solar wind's influence extends all the way to the surface. However, the good agreement of the $u_i=u_n$ model at low activity (Eq. \ref{eq: FFCF}) with the plasma density measurements is seemingly at odds with this.

Putting together this complex picture of the ion dynamics requires the ions to be treated kinetically, as is done in test-particle model and hybrid plasma simulations. Furthermore, detailed treatment of the ion-neutral collisions (i.e., to assess the extent of collisional decoupling) is key. This study aims to shed light on the effect of a more-complex 3D solar wind--comet interaction on the ionospheric composition and density, with a view to better interpret the RPC data for intermediate outgassing, between low outgassing conditions (Eq.~\ref{eq: FFCF})  and the formation of a diamagnetic cavity \citep{Lewis2024}.

\subsection{Structure of the paper}

For this study, the collisional test-particle model originally developed for the electron populations around a comet \citep{Stephenson2022} has been adapted to instead describe the ion environment. We model the three key cometary ion species ($\mathrm{H_2O^+}$, $\mathrm{H_3O^+}$, and $\mathrm{NH_4^+}$) and the collisions between them and the neutral coma background. In Section~\ref{sec: test pl description}, the model is described and validated, with emphasis on the aspects of the code that were updated from the electron model. Section~\ref{sec: model collisions} describes the treatment of ion-neutral collisions in the model, including a comparison of the 3D kinetic approach and the 1D fluid approach to the diamagnetic cavity case presented in \citet{Lewis2024}. Section~\ref{section: hybrid fields} then describes the hybrid simulations that provide the electric and magnetic fields to drive the ions in the test-particle model. 

Section~\ref{sec: application test pl} focuses on the cometary plasma environment around 67P at $2.5-3$~au, when the outgassing was lower and the diamagnetic cavity was not detected.  The density and bulk velocity of $\mathrm{H_2O^+}$, $\mathrm{H_3O^+}$, and $\mathrm{NH_4^+}$ are assessed, highlighting the extent to which each is driven by the changing ion transport and collisions.

Finally, in Section~\ref{sec: chap 5 data comparison}, the combined dataset from MIP/LAP is compared with the results of the test-particle modelling in Section~\ref{sec: application test pl}. The aim of this is to understand the drivers of the plasma density during the conditions used for the hybrid simulations, and to assess how well the simulation recreates them. 

\section{Modelling} 
\label{sec: modelling}
\subsection{Collisional ion test-particle model}
\label{sec: test pl description}
 The test-particle model first described in \citet{Stephenson2022} was developed to assess the response of cometary and solar wind electron populations to electric and magnetic fields provided by a fully kinetic Particle-in-Cell (PiC) model (\cite{Deca2017}, \cite{Deca2019}). Crucially, the test-particle approach enabled study of the effect of collisions on the overall electron environment, which have not yet been incorporated into kinetic PiC models (\cite{Stephenson2023a}, \cite{Stephenson2024}). 

\begin{figure*}
    \centering
    \includegraphics[width=0.8\linewidth]{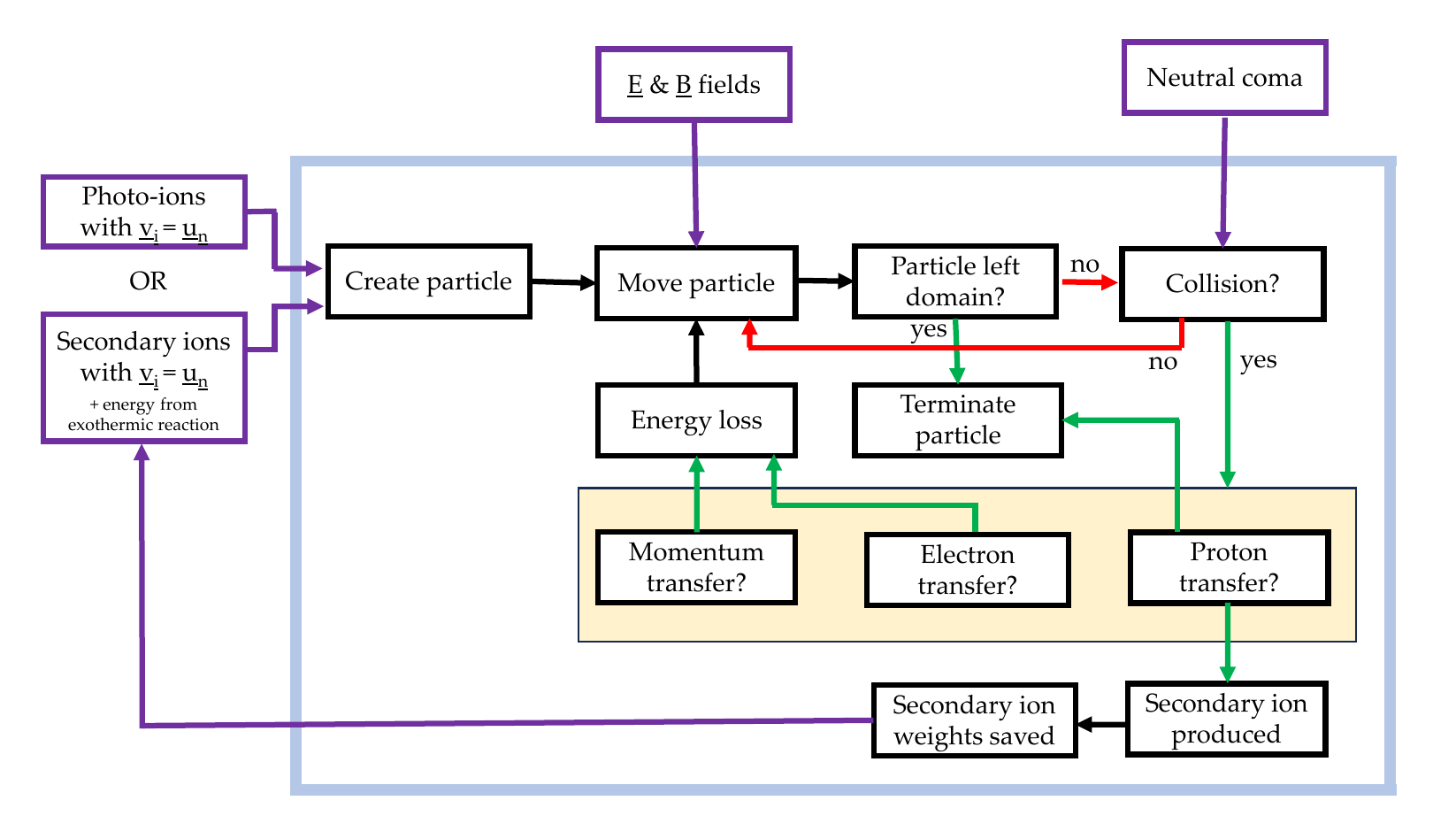}
    \caption{Flow chart representing the key elements of the ion test-particle model. The {blue} box separates processes happening inside the code, and the inputs controlled externally (in {purple boxes}). Adapted from \citet{Stephenson2022} for the application to cometary ions.}
    \label{fig: test-particle flow chart}
\end{figure*}

In this study, we have adapted the \citet{Stephenson2022} electron test-particle model to now model the cometary ion population. The model is described by the flow chart in Figure~\ref{fig: test-particle flow chart}. First, the ions are created, either as photo-ions ($\mathrm{H_2O^+}$) or as secondary ions ($\mathrm{H_3O^+}$, $\mathrm{NH_4^+}$), produced through ion-neutral chemical reactions. $\mathrm{H_2O^+}$ ions are initialised in the model with a given velocity and position in the grid. Newly produced ions originate from the neutral coma, they are initially given the same velocity as the neutrals: $\vec{u_i} = u_n \hat{r}$. The total unattenuated ionisation frequency $\nu^{\mathrm{tot}}_{ioni}$ is supplied as an input to the model and subsequently calculated in 3D taking photo-absorption into account, using the Lambert-Beer law.

The simulated ions (macroparticles) represent a flow of ions that follow the same trajectory through the fields. $\mathrm{H_2O^+}$ macroparticles are assigned a `weight' ($W_p$), based on the number of ions per second which follow the same path. This weight is dependent on the number of other ions produced within the same grid cell. It is given by 

\begin{equation}
    W_p = \frac{1}{N_p}\int_{\text{cell}} \nu_\mathrm{ioni}^{\mathrm{tot}}(\vec{x}) n_n(\vec{x}) d^3\vec{x}
\end{equation}

where $N_p$ is the number of macroparticles produced within the cell and $n_n(\vec{x})$ is the neutral density at the position $\vec{x}$ within the grid. The inner part of the production grid comprises of 20 small boxes ($1~\mathrm{km} \times 1~\mathrm{km} \times 1~\mathrm{km}) $, extending from the origin, where the comet nucleus is centred. The rest of the ($1000~\mathrm{km} \times 1000~\mathrm{km} \times 1000~\mathrm{km}) $ domain is then filled with logarithmically spaced boxes such that there are 50 cells in the x-direction (the comet-Sun line) and 40 in the y- and z-directions.

The particles are then pushed through the background fields by the Lorentz force. At each timestep, we check whether a collision has occurred, and if so, what type. Particles are terminated when they either leave the domain, hit the nucleus, or undergo a proton transfer. After the run for the photo-ions ($\mathrm{H_2O^+}$) has been completed, the secondary ions are generated from the weights that have been saved when proton transfer collisions occur. 

Finally, the moments of the distributions can be calculated based on the trajectories of the particles, provided enough particles have travelled through each cell. For details of the moment calculation, see \citet{Stephenson2022}. The moments are calculated on a $(80\times 60 \times 60)$ cell grid with the same resolution as the electric and magnetic fields (25~km).

\subsection{Hybrid electric and magnetic fields}
\label{section: hybrid fields}

The electric and magnetic fields are provided by 
 AMITIS (Advanced Modelling InfrasTructure in Space simulations, \cite{Fatemi2017}), a GPU-based 3D hybrid simulation that has been applied to various solar system bodies. The same run as presented in \citet{Moeslinger2024} is used. The ion production rate used as input to the AMITIS model run is $Q\nu = 1.08 \times 10^{20}~\mathrm{s^{-2}}$, and corresponding $Q$ and $\nu$ values for the test-particle model input are determined from the COPS data and total ionisation rates \citep{Stephenson2023a}. The model inputs are summarised in Table~\ref{tab: Amitis inputs}, and are broadly representative of the conditions at 67P encountered by Rosetta at $2.5-3$~au. 

\begin{figure*}
    \centering
    \includegraphics[width=0.8\linewidth]{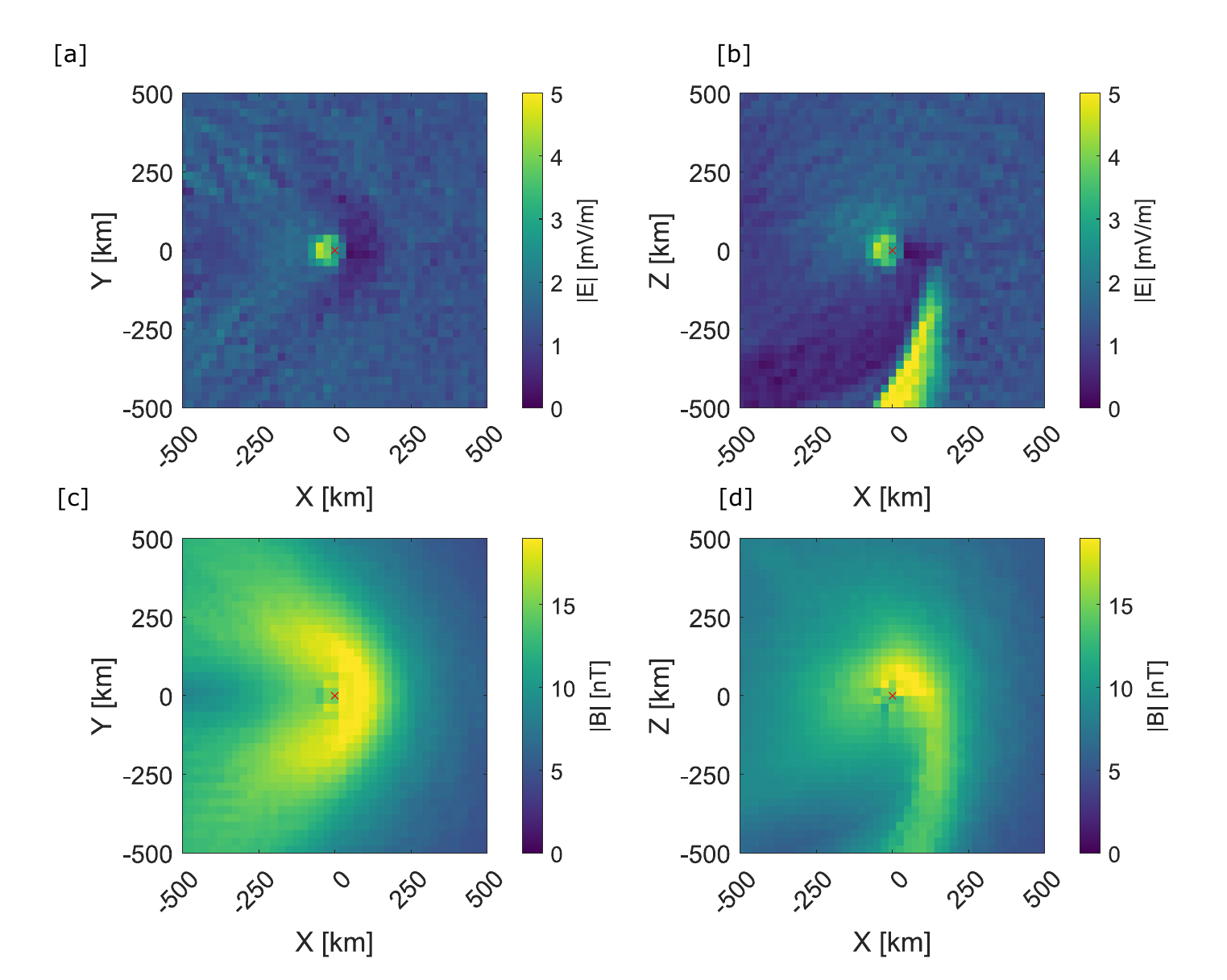}
    \caption{{[a]} and {[b]} Electric field and {[c]} and {[d]} magnetic field magnitudes from AMITIS \citep{Fatemi2017}, used to drive the collisional test-particle model \citep{Moeslinger2024}. Red cross inidicates the position of the cometary nucleus.}
    \label{fig: amitis fields}
\end{figure*}

\begin{table}
	\centering
	\caption{Inputs to the AMITIS hybrid simulation as used in Sections~\ref{sec: application test pl} and \ref{sec: chap 5 data comparison}. The same inputs (where relevant) were applied to the test-particle model, except the moments were calculated over a $1000\times1000\times 1000$~km grid.}
	\label{tab: Amitis inputs}
	\begin{tabular}{l|c} 
         \hline

       $Q ~ \mathrm{[s^{-1}]}$  & $5.4 \times 10^{26}$    \\
       $\nu~ \mathrm{[s^{-1}]}$ &  $2 \times 10^{-7}$\\ 
      $u_n~ \;[\mathrm{km~s^{-1}}]$ & $0.7$ \\ 
       Simulation domain [km] & $7000 \times 12000 \times 16000$ \\ 
       Grid resolution [km]  & $25$ \\ 
       Heliocentric distance [au] & $2.5 - 3$ \\ 
       Upstream $n_{\mathrm{SW}}$ [$\mathrm{cm^{-3}}$] &  1 \\ 
       Upstream $u_{\mathrm{SW}}~[\mathrm{km~s^{-1}}]$ & 430 \\ 
       Upstream $T_{e,\mathrm{sw}}$~ [$\mathrm{eV}$] & 12 \\ 
       Upstream $B_{\mathrm{IMF}}$~ [$\mathrm{nT}$] & 3 \\	\hline
	\end{tabular}
    \label{tab: hybrid comparisons}
\end{table}

The electric and magnetic field magnitudes are showin in Figure~\ref{fig: amitis fields}. The Sun in the +X direction, and the interplanetary magnetic field is oriented along the +Y axis. The validation of the ion test-particle model under collisionless conditions is shown in Appendix~\ref{sec: collisionless validation appendix}, using the input parameters given in Table~\ref{tab: Amitis inputs}.

\subsection{Ion-neutral collisions in 3D}
\label{sec: model collisions}
Adapting the test-particle model to treat the cometary ion population requires a new set of cross sections. The full ion-neutral cross section set and references are presented in Appendix~\ref{sec: test particle cross secs appendix}. In general, the cross sections increase with decreasing relative energy, but the experimental cross sections are only available down to 0.1 eV. For relative energies below this value, values were linearly extrapolated in log space down to an approximate surface temperature of the neutrals of $\sim 0.01 \; \mathrm{eV}$ (\cite{Heritier2017a}, \cite{Marshall2017}).

 Dissociative recombination is neglected, which is justified at the outgassing rates used in this study ($10^{26}-10^{27} \; \mathrm{s^{-1}}$, reflecting January--March 2016 during Rosetta). \citet{Heritier2018a} demonstrated that neglecting recombination in this regime introduces less than $20\%$ error (for $T_e = 200\;\mathrm{K}$) in the total plasma density. This is even lower when acceleration of the ions by electromagnetic fields is taken into account (see \citet{Lewis2024}). 

 The probability of a given collision type $c$ happening during a timestep $\mathrm{d}t$ is given by:

\begin{equation}
    P_{c}(\mathrm{d}t) = 1 - \exp \left [-n_n(\vec{x}) \: \mathrm{d}V_{\sigma} \right ]
\end{equation}
where $\mathrm{d}V_{\sigma}$ is the volume `swept through' by the collision cross section in the frame of the neutral gas (of relevant species, $n$). This volume is given by
\begin{equation}
    \mathrm{d}V_{\sigma} = \sigma_c(E_\mathrm{rel}) \;v_{\mathrm{rel}} \; \mathrm{d}t
\end{equation}
where $v_\mathrm{rel}$ is the relative speed between an ion and a neutral and the collision cross section is $\sigma_c(E_\mathrm{rel})$. 

For a neutral gas with a Maxwell velocity distribution function $f_n(\vec{v_n})$, a bulk velocity $\vec{u}_n$, and a temperature $T_n$, the relative speed is defined as \citep{Fahr1967}: 
\begin{equation}
    v_\mathrm{rel} = \iiint |\vec{v_i} - \vec{v}_n| \; f_n(\vec{v_n}) \; \mathrm{d}^3\vec{v_n}. 
\label{eq: gross erf thing}
\end{equation}
Eq.~\ref{eq: gross erf thing} can be approximated as
\begin{equation}
    v_\mathrm{rel} \approx \left ( \frac{4 v_{n,th}^2}{\pi} + w_\mathrm{rel}^2 \right)^\frac{1}{2} = \left ( (v_{n,th}^{\text{mean}})^2 + w_\mathrm{rel}^2 \right )^{1/2}
    \label{eq: rel vel update}
\end{equation}

where $v_{n,th} = \sqrt{\frac{2k_BT_n}{m_n}}$ is the location of the peak of the Maxwellian velocity distribution (in the frame of the gas), that is, the most probable speed. $v_\mathrm{rel}$ is a function of both the difference between the bulk velocities, $w_{\mathrm{rel}} = |\vec{v_i}(\vec{x}) - \vec{u_n}(\vec{x})|$, and the thermal speed, $v_{n,th}^{\text{mean}}$, corresponding to the mean speed in the frame of the neutral gas. Eq.~\ref{eq: rel vel update} is equivalent within $5 \%$ over all $w_\mathrm{rel}$ to the full solution to Eq.~\ref{eq: gross erf thing}. $v_{\mathrm{rel}}$ can then be used to calculate the relative energy $E_{\mathrm{rel}}$ in order to find the cross section $\sigma_c(E_\mathrm{rel})$ using
\begin{equation}
    E_\mathrm{rel} = \frac{1}{2} \frac{m_j m_n}{m_j + m_n} v_{\mathrm{rel}}^2.
\end{equation}

The 3D model allows us to account for the exothermic energy release from the protonation reaction. In this case, a component of velocity with a magnitude equivalent to the $0.5~\mathrm{eV}$ energy release \citep{Hunter1998} is given to secondary ions (those newly produced through protonation) in a randomly generated direction. 

The model has been validated using the energy-independent kinetic rates as used in the 1D ion acceleration model of \citet{Lewis2024}, with the inputs given in Table~\ref{tab: validation parameters}. The validation is shown in Appendix~\ref{sec: radial fields validation appendix}.

\subsubsection{1D fluid vs 3D kinetic modelling: diamagnetic cavity}
\label{sec: new colls dm cavity}

The 3D kinetic approach can be used to assess the validity of the 1D fluid approach, applied within the diamagnetic cavity by \citet{Lewis2024}. To do so, the hybrid electric and magnetic field inputs are replaced with $\vec{B}=\vec{0}$ everywhere in the simulation, and a radial electric field which is given by $\vec{E} = (E_cr_c/r)  ~\hat{r}$, with $E_c = 0.5 \times 10^{-3}~\mathrm{Vm^{-1}}$ the field strength at the comet surface. The electric field is calculated over a $(1000 \times 1000 \times 1000) ~\mathrm{km}$ grid with a $10~\mathrm{km}$ spatial resolution in each direction. Since there is no magnetic field in this case, only a radial electric field, the set-up is valid only inside the diamagnetic cavity, and represents inputs similar to those seen on 21st November 2015 by Rosetta, illustrated in Table~\ref{tab: validation parameters}.

\begin{table}
    \centering
    \caption{Inputs used to compare the 1D ion acceleration model and present test-particle model in Section~\ref{sec: new colls dm cavity}. Inputs are data-based and justified in \citet{Lewis2024}.}
    \begin{tabular}{l|l}
         Rosetta equivalent date & 2015-11-21 \\ \hline
         $Q \; [\mathrm{s}^{-1}]$ & $7 \times 10^{27}$ \\
         $\nu \; [\mathrm{s}^{-1}]$ & $2 \times 10^{-7}$ \\
         $\% \mathrm{NH_3}$ & 0.2 \\
         $ u_n \; [\mathrm{km \;s^{-1}}]$ & 0.75 \\
         $E(r_c) \; [\mathrm{mV \; m^{-1}}]$ & 0.5 
    \end{tabular}
    \label{tab: validation parameters}
\end{table}

\begin{figure*}
    \centering
    \includegraphics[width=\linewidth]{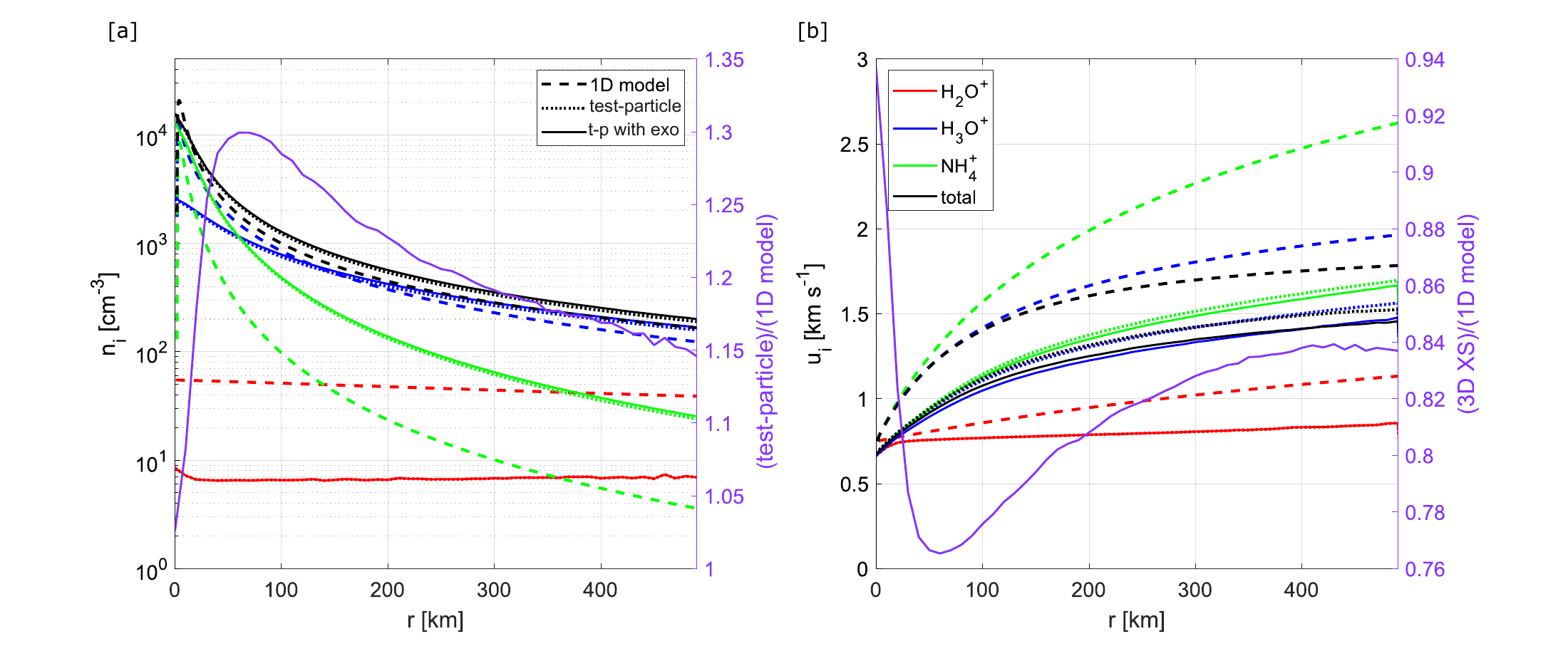}
    \caption{Comparison of 1D Ion Acceleration Model from \citet{Lewis2024} ({dashed lines}) with the 3D ion test-particle model with ({solid lines}) and without ({dotted line}) an exothermic energy release during protonation for [a] ion density and [b] ion bulk velocity. Profiles for $\mathrm{H_2O^+}$ ({red}), $\mathrm{H_3O^+}$ ({blue}) and $\mathrm{NH_4^+}$ ({green}) are shown, alongside the total for all species ({black}). Right-hand axis shows the ratio between the two models for the total ion density ({purple}). }
    \label{fig: compare new colls}
\end{figure*}

The implication of these energy-dependent cross sections on the ion composition and velocity profiles is shown in Figure~\ref{fig: compare new colls}. Overall, the main effect of the 3D treatment is to boost the proton transfer, reducing the $\mathrm{H_2O^+}$ density and producing more $\mathrm{NH_4^+}$. This is because the energy-dependent cross sections peak at low energies. The total plasma density is increased by a factor of up to 1.3 in the 3D model thanks to more effective collisions, and the ion bulk velocity is up to 23\% slower.  However, as demonstrated in \citet{Lewis2024}, the product $n_iu_i$ for a given $r$ is preserved, so the determination of the electric field strength and bulk velocity that are required to recreate the measured plasma density from RPC--LAP and RPC--MIP remains valid. 

Figure~\ref{fig: compare new colls} also shows effect of an $0.5~\mathrm{eV}$ exothermic energy release during protonation (solid line). The difference is very small (in some cases not visible on the plots), only reducing the total plasma density by a maximum of $5\%$, and increasing the bulk velocity by up to $8\%$. 

\section{Cometary ion dynamics at 2.5-3 au}
\label{sec: application test pl}

We now run the full collisional ion test-particle simulation, driven by the hybrid electric and magnetic fields (input parameters in Table~\ref{tab: Amitis inputs}, see Section~\ref{section: hybrid fields}), representing comet 67P at 2.5-3~au. The neutral coma is assumed to be $\mathrm{H_2O}$-dominated, with a $0.2~\%$ contribution of $\mathrm{NH_3}$. The three ion species are $\mathrm{H_2O^+}$, $\mathrm{H_3O^+}$, and $\mathrm{NH_4^+}$ (see Section~\ref{sec: test pl description}).

\begin{figure*}
    \centering
    \includegraphics[width=0.9\linewidth]{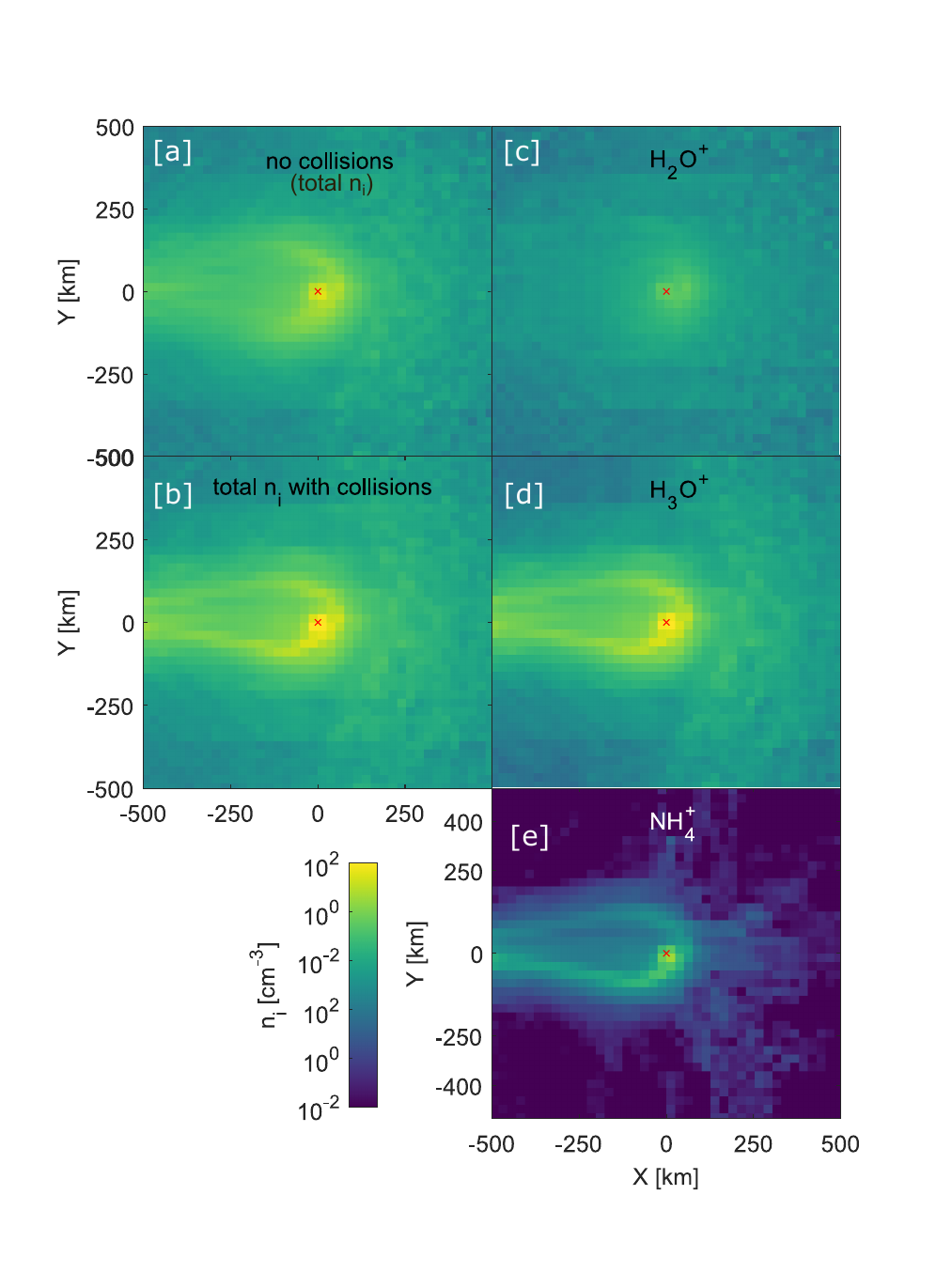}
    \caption{Ion density in the X-Y plane from the ion test-particle model, using the electric and magnetic fields from the AMITIS hybrid simulation (Table \ref{tab: hybrid comparisons}). {[a]} total ion density with no collisions, [b] total ion density with all collisions. For the collisional case, densities of {[c]} $\mathrm{H_2O^+}$, {[d]} $\mathrm{H_3O^+}$, and {[e]} $\mathrm{NH_4^+}$ are given. The Sun is in the +X direction, and the comet nucleus is marked with a red cross.}
    \label{fig: amitis density}
\end{figure*}

Figure~\ref{fig: amitis density} shows the modelled ion density in the X-Y plane, with the Sun in the +X direction. Comparing Figure~\ref{fig: amitis density}a (no collisions) and Figure~\ref{fig: amitis density}b (collisions), the inclusion of collisions increases the total ion density, and the ions are slowed down. 

\begin{figure}
    \centering
    \includegraphics[width=\linewidth]{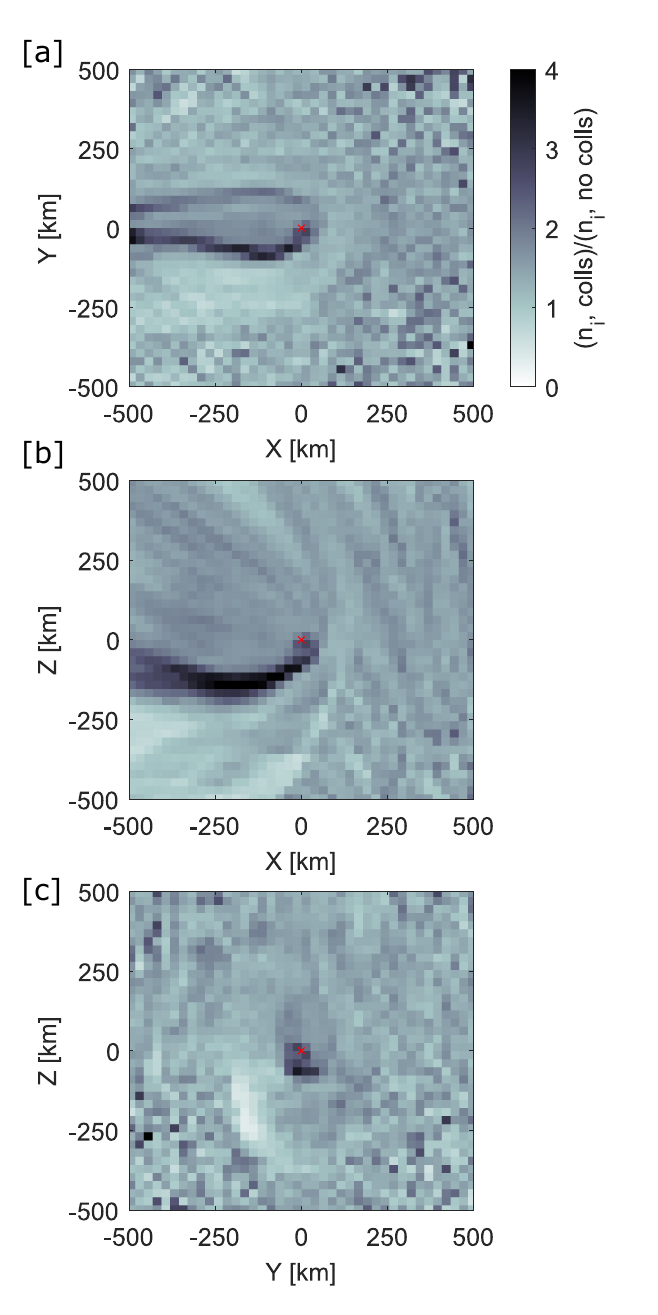}
    \caption{Ratio between the total ion densities calculated from the test-particle model with collisions included (as in Fig. \ref{fig: amitis density}b) and without (as in Fig. \ref{fig: amitis density}a).{ Red cross} marks the location of the nucleus.}
    \label{fig: amitis compare colls}
\end{figure}

The ion density is largest close to the nucleus, as in the 1D case, but it is also enhanced just inside the region of magnetic pile up (see Figure~\ref{fig: amitis fields}c). The cometary ions appear to stagnate in this area, which was also observed in the collisionless simulations of \citet{Moeslinger2024}, but the addition of ion-neutral collisions increases the piled-up density. Figure~\ref{fig: amitis compare colls} shows the ratio between the total ion density in the collisional and non-collisional case, in all three planes. Collisions are shown to increase the total ion density by up to 4 times, with the greatest impact on the densities in the region of ion pile up, on the inner edge of the magnetic pile up region. It can also be seen in Figure~\ref{fig: amitis compare colls}b that the effect of collisions is more significant for -Z values. This is likely due to enhanced transport in the +Z direction by the motional electric field. 

When the ion density is separated into species (see Figure~\ref{fig: amitis density}c-e), there is a clear difference in behaviour of the three key ion species. As ionisation and protonation processes are predominant in the inner coma, all three ion species are produced near the nucleus with a similar velocity to the neutral gas. However, all species are affected differently by the electric field. This is not simply because of different electromagnetic forces on them, since they all have the same charge ($+ e$) and very similar masses ($18-19 \; \mathrm{u \; q^{-1}}$). Instead, the difference in their ion-neutral chemical timescales drives this variation. 

$\mathrm{H_2O^+}$ (Figure~\ref{fig: amitis density}c) is around an order of magnitude less prevalent and significantly more homogenous than $\mathrm{H_3O^+}$ (Figure~\ref{fig: amitis density}d), with no clear enhancement in the pile up region. $\mathrm{H_2O^+}$ reacts very quickly with $\mathrm{H_2O}$ to form $\mathrm{H_3O^+}$, therefore it has too little time to undergo significant acceleration by the electric field. $\mathrm{H_2O^+}$ is at (or close to) photochemical equilibrium in regions close to the nucleus, and directly lost through transport far from the nucleus. 

$\mathrm{NH_4^+}$ (Figure~\ref{fig: amitis density}e) is produced in low densities (with similar magnitudes to $\mathrm{H_2O^+}$) at its peak, but is concentrated more strongly near the nucleus and in the ion pile up region. This is because $\mathrm{NH_4^+}$ is particularly sensitive to the increased ion transport further from the nucleus, and doesn't have time to be produced from ion-neutral chemistry when the bulk flow is too fast. $\mathrm{NH_4^+}$ is also more strongly accelerated and reached higher velocities. 

\begin{figure*}
    \centering
    \includegraphics[width=0.9\linewidth]{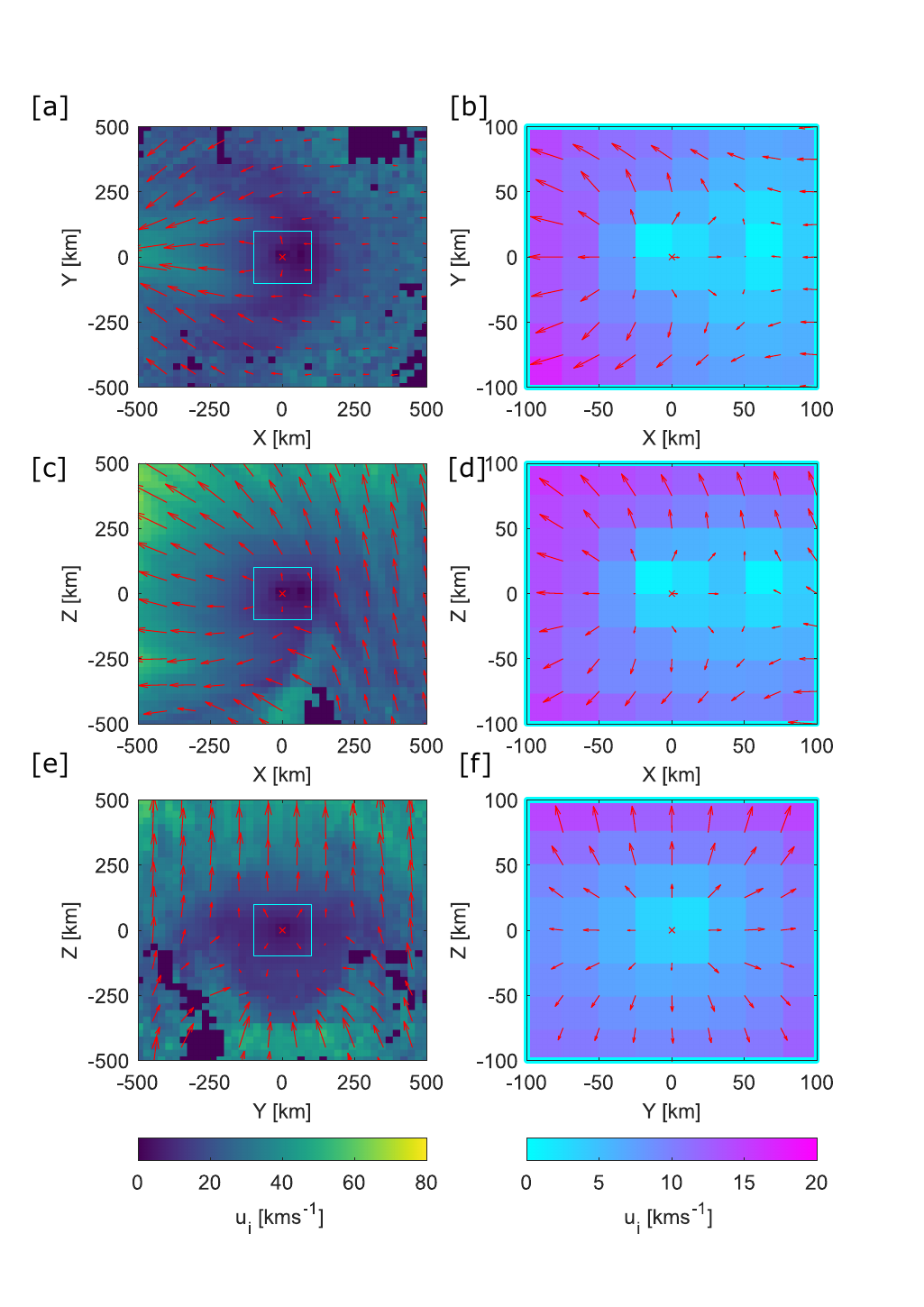}
    \caption{Ion bulk velocity magnitude ({colour scale}) and direction ({red arrows}) in the {[a, b]} X-Y plane, {[c, d]} X-Z plane, and {[e, f]} Y-Z plane, generating using the ion test-particle model with collisions included. {Cyan box }in left column corresponds to the $100 \times 100$ inset, shown in the right column. The comet nucleus is marked with a {red cross}.}
    \label{fig: amitis quivers}
\end{figure*}

Figure~\ref{fig: amitis quivers} shows the total ion bulk velocity with arrows to show the direction of travel in each plane. Collisions are included for this simulation. Figures \ref{fig: amitis quivers} [b], [d], and [f] show the same information as [a], [c] and [e], but focussed on the first $100~\mathrm{km}$ from the nucleus. Panel [a] shows the bulk flow mainly diverted around the nucleus and tail-ward, with the radial motion only visible for $0 - 50\;\mathrm{km}$ from the nucleus. In the Y-Z plane (Figure~\ref{fig: amitis quivers}f), where Rosetta was usually located, the radial flow extends over a larger region, up to 100~km. Figures \ref{fig: amitis quivers} [c] and [e] show a strong component of the flow in the + Z direction, owing to the motional electric field.

\section{Comparison with Rosetta data}
\label{sec: chap 5 data comparison}
In this section, the modelled ion density is compared to the data from RPC--MIP and RPC--LAP. First, the full RPC--MIP/LAP combined dataset \citep{Johansson2021} compared against the plasma density calculated by the field-free, chemistry-free model (Eq.~\ref{eq: FFCF}) to verify where the model performs best over the Rosetta escort phase. This provides a reference case to compare with the ion test-particle model, to assess whether the plasma density can be constrained more accurately by the more complex model. In Section ~\ref{sec: MIP/LAP vs test particle}, the ion test-particle model is compared against the MIP/LAP dataset and the field-free, chemistry-free model. 

\subsection{MIP/LAP data vs field-free chemistry-free model}

\begin{figure*}
     \centering
     \includegraphics[width=\linewidth]{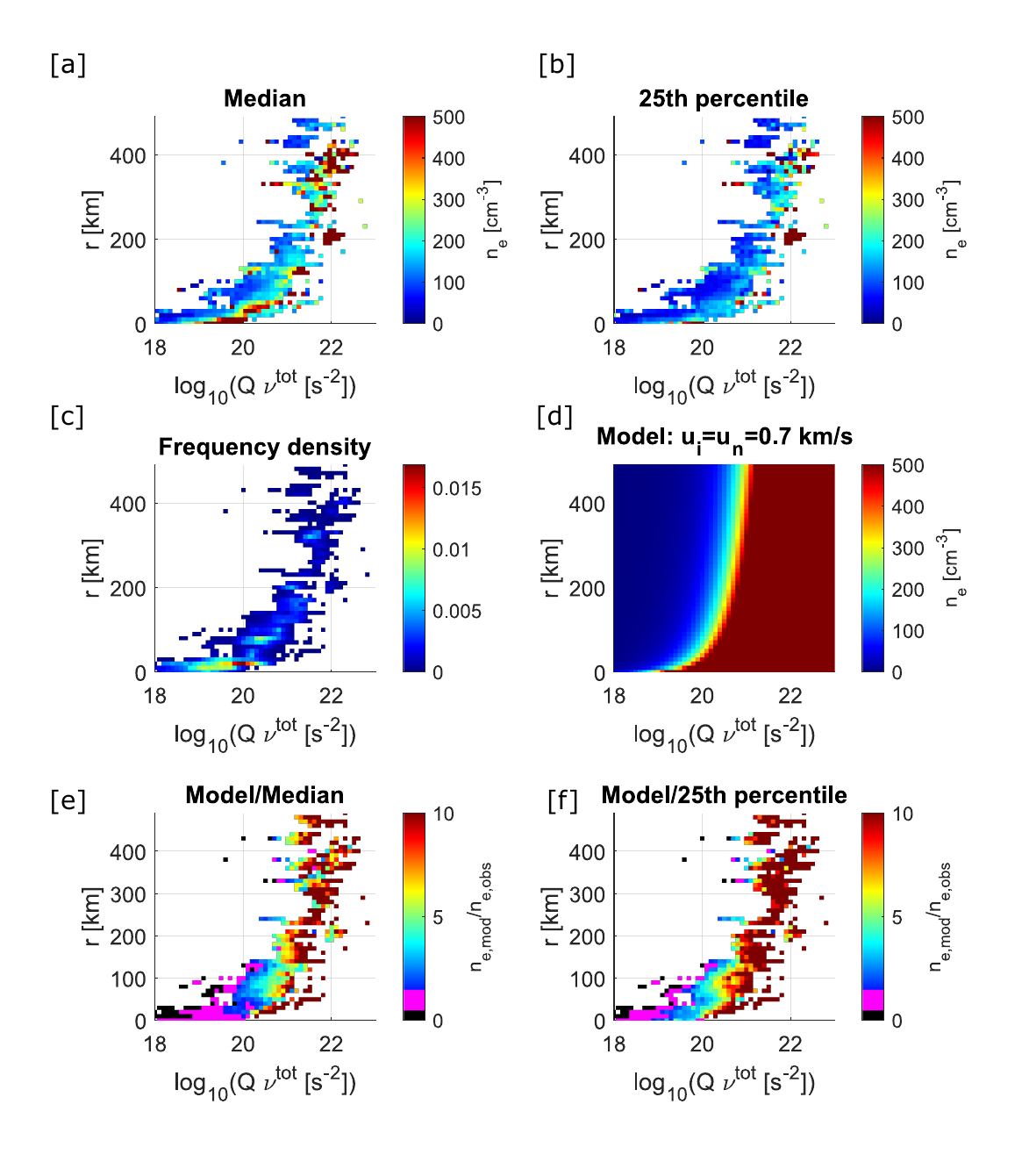}
     \caption{Heat maps showing the MIP/LAP combined dataset from the whole escort phase (covering August 2014-- August 2016), binned by cometocentric distance $r$ and $Q \nu^{\mathrm{tot}}$. {[a]} Shows the median electron density in each bin, {[b]} shows the 25th percentile. {[c]} shows, for context, the frequency density of data in each bin. {[d]} Shows the density as calculated by the 1D field-free model (Eq.~\ref{eq: FFCF2}) and {[e]} and {[f] }the ratio between the field-free model and the MIP/LAP median and 25th percentiles respectively, with {magenta} indicating where the ratio is 0.5--1.5.}
     \label{fig: ne dens}
 \end{figure*}

The MIP/LAP electron density data \citep{Johansson2021} are first smoothed with a 1-minute rolling median to reduce noise and small-scale variations. Figure~\ref{fig: ne dens} shows the smoothed dataset sorted into $10\;\mathrm{km}$ bins of cometocentric distance and bins of width 0.1 of $\log_{10}{(Q\nu^{\mathrm{tot}})}$. 
The median of each bin (Figure~\ref{fig: ne dens}a) is shown, as well as the 25th percentile, as this value is less sensitive to electron density spikes, and may capture the baseline plasma density better. $Q \nu^{\mathrm{tot}}$~[$\mathrm{s^{-2}}$] is calculated from the local outgassing $Q$ (from COPS and corrected for the neutral composition) and the total $\mathrm{H_2O}$ ionisation frequency (photoionisation and electron impact) from \citet{Stephenson2023a}. 

Figure~\ref{fig: ne dens}d then shows the field-free, chemistry-free model,

\begin{equation}
    n_i(r) = \frac{Q\nu^{\mathrm{tot}}}{4\pi u_n^2 r^2} (r-r_c),
    \label{eq: FFCF2}
\end{equation}
which is equivalent to Eq.~\ref{eq: FFCF}. The bulk neutral velocity $u_n$ is known to vary with heliocentric distance \citep{Biver2019} as well as with cometocentric distance \citep{Heritier2017a}, but is assumed constant for Eq.~\ref{eq: FFCF2}. A value of $u_n = 700~\mathrm{km~s}^{-1}$ is assumed, in line with the inputs to the ion test-particle model and hybrid simulation (Table~\ref{tab: hybrid comparisons}). The colour scale in Figure~\ref{fig: ne dens}d is capped at $500~\mathrm{cm^{-3}}$.

Figure~\ref{fig: ne dens}c highlights the limitations of the parameter space we have for the Rosetta data. Despite its extensive escort phase, practical limitations mean that the coverage for low cometocentric distances at high outgassing is limited, and vice versa. However, it is possible to draw several conclusions from the data we do have. Figures \ref{fig: ne dens}e and \ref{fig: ne dens}f show the overestimation of the field-free chemistry-free model (Eq.~\ref{eq: FFCF2}) increases with ion production rate. At low outgassing, as predicted by \citet{Galand2016} and demonstrated in \citet{Galand2016} and \citet{Heritier2018a}, the field-free model performs well to explain the measured electron density (magenta). At high outgassing, where the conditions of the field-free chemistry-free model break, the observed electron density is up to 40 times higher than is calculated (colour scale in Figure~\ref{fig: ne dens}e and \ref{fig: ne dens}f are capped at 10 for clarity). This has been previously reported in e.g. \citet{Vigren2019}, but the full comparison over the whole escort phase is made here for the first time, using the full ionisation rates (photoionisation and electron-impact ionisation) as provided by \citet{Stephenson2023a}. Where the data are higher than the field-free chemistry-free model (shown in black in Figure~\ref{fig: ne dens}e and f), this may be due to the changes in neutral composition at low outgassing. The $\mathrm{CO_2}$ component of the coma was found to increase significantly over the southern hemisphere beyond 3.1~au \citep{Gasc2017}, which was shown in \citet{Galand2016} to increase the photoionisation rate by up to 50\%.

\begin{figure*}
    \centering
    \includegraphics[width=\linewidth]{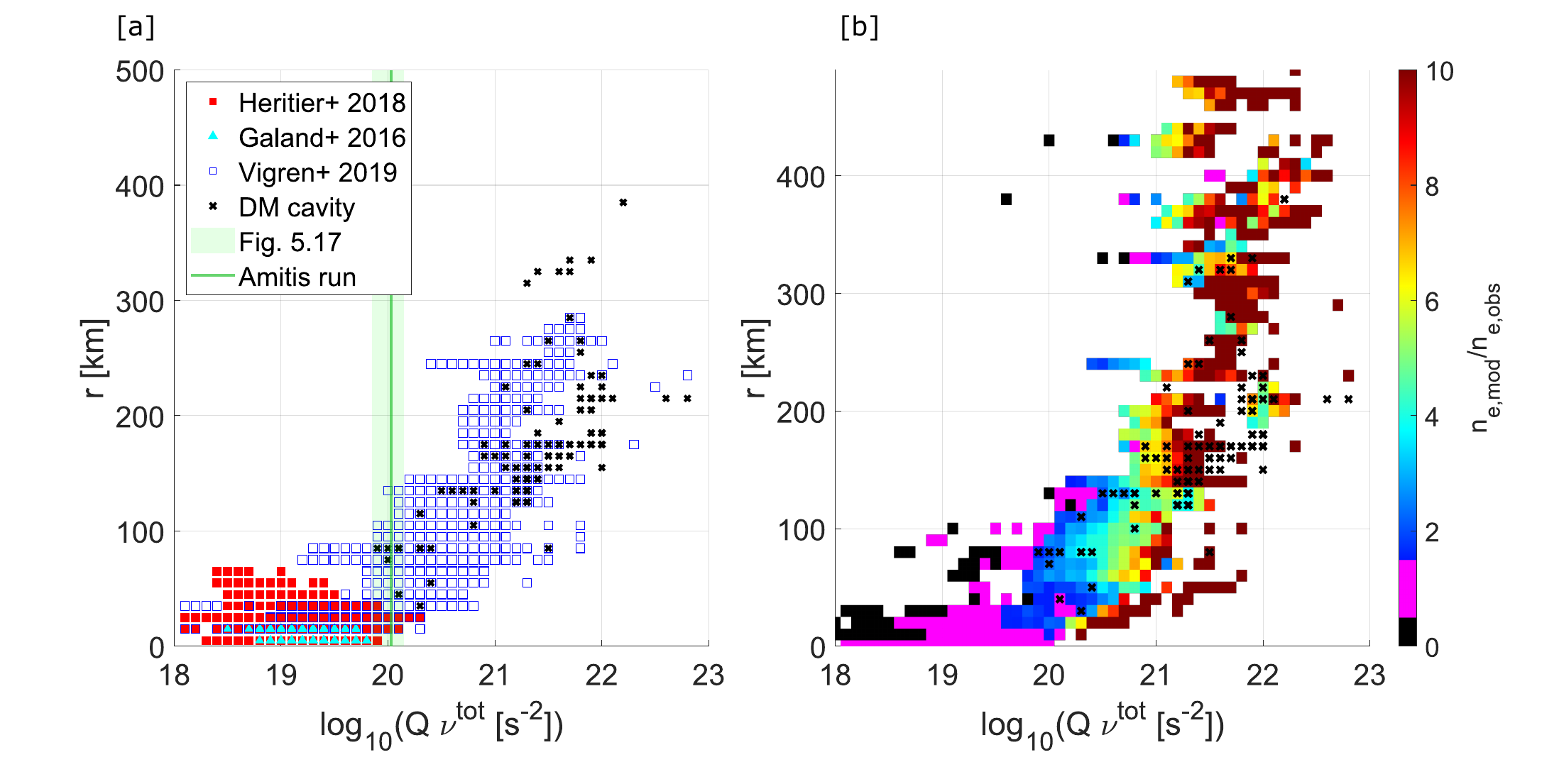}
    \caption{({a}) Coverage of the parameter space in $Q~\nu$ vs $r$ (as in Figure~\ref{fig: ne dens}), done by previous studies: \protect\cite{Heritier2018a} ({red filled squares}),  \protect\cite{Galand2016} ({cyan triangles}), and \protect\cite{Vigren2019} ({blue squares}). The ion production rate $Q~\nu$ from the hybrid simulation run is given by the {green line}, and the {green shaded region} is the data range used for MIP/LAP in Figure~\ref{fig: ne dens}. The {black crosses} show diamagnetic cavity crossings as observed by RPC--MAG \protect\citep{Goetz2016a} ({b}) Same as Figure~\ref{fig: ne dens}e but with the diamagnetic cavity crossings marked with {black crosses}.}
    \label{fig: parameter space}
\end{figure*}

The coverage of the parameter space in Figure~\ref{fig: ne dens} by previous studies is shown in Figure~\ref{fig: parameter space}a, alongside the ion production rate $Q~\nu^{\mathrm{tot}}$ used in this study (green line), and the instances of diamagnetic cavity crossings as detected by RPC--MAG \citep{Goetz2016a} (black crosses). Both the studies of \citet{Galand2016} (cyan triangles) and \citet{Heritier2018a} (red squares) reported a good agreement between the field-free, chemistry-free model (Eq.~\ref{eq: FFCF2}) and the electron density data\footnote{Note that MIP and LAP were used separately in these studies as the combined MIP/LAP dataset \citep{Johansson2021} was not yet available. This means that these studies include times when only one of the two datasets was available, which are not included in the combined dataset.} in the low outgassing cases (for which the model was developed). The study of \citet{Vigren2019} (blue squares) applied Eq.~\ref{eq: FFCF2} over a large time range post-perihelion, showing a transition from overestimation of the model near perihelion to good agreement at lower ion production rates. This result is well replicated in Figure~\ref{fig: ne dens}, and generalised over the full dataset.

Figure~\ref{fig: parameter space}b shows the location of the diamagnetic cavity boundary crossings as seen by RPC--MAG overlaid on Figure~\ref{fig: ne dens}e. In general, the diamagnetic cavity has been observed for $Q \nu^{\mathrm{tot}}>1\times 10^{20}~\mathrm{s^{-2}}$, where Eq.~\ref{eq: FFCF2} overestimates the plasma density. The larger the ion production rate, the further from the nucleus the cavity has been observed - however the data bias in the parameter space explored by Rosetta is probably artificially enhancing this trend. 

 The ion production rate $Q\nu^{\mathrm{tot}} = 1.08\times10^{20}$~s$^{-2}$ used in the simulation in the hybrid simulation falls in an interesting region of the parameter space shown in Figure~\ref{fig: parameter space}a. At cometocentric distances below 50~km, a generally good agreement with the field-free, chemistry-free model has been reported (\cite{Heritier2018a} and in Figure~\ref{fig: ne dens}). However, diamagnetic cavity crossings have also been observed at other times when Rosetta was closer to 100~km for the same $Q~\nu^{\mathrm{tot}}$. It is possible that this ion production rate reflects somewhat of a transition between the low outgassing regime, where the field-free chemistry-free model applies, and where the ambipolar electric field becomes important and leads to increased transport at the Rosetta distance. 

\subsection{MIP/LAP data vs test-particle model at 2.5-3~au}
\label{sec: MIP/LAP vs test particle}

The plasma density from the ion test-particle model is now compared with the MIP/LAP combined dataset and the field-free chemistry-free model. The ion test-particle model and the AMITIS hybrid simulation only treat photoionisation, and do not include the electron-impact ionisation frequency. To account for this shortcoming of the models, both have artificially boosted the photoionisation frequency in accordance with the electron-impact ionisation frequency measured by RPC--IES (e.g., \cite{Stephenson2023a}). The input parameters in Table \ref{tab: hybrid comparisons} for the AMITIS model then represent the total ionisation frequency ($\nu^{\mathrm{tot}}$) for the $2.5 - 3.0$~au range. This total value is used in this Section for comparison with the Rosetta data. A more thorough implementation of electron-impact ionisation into the model is left for future work.

\begin{figure*}
    \centering
    \includegraphics[width=\linewidth]{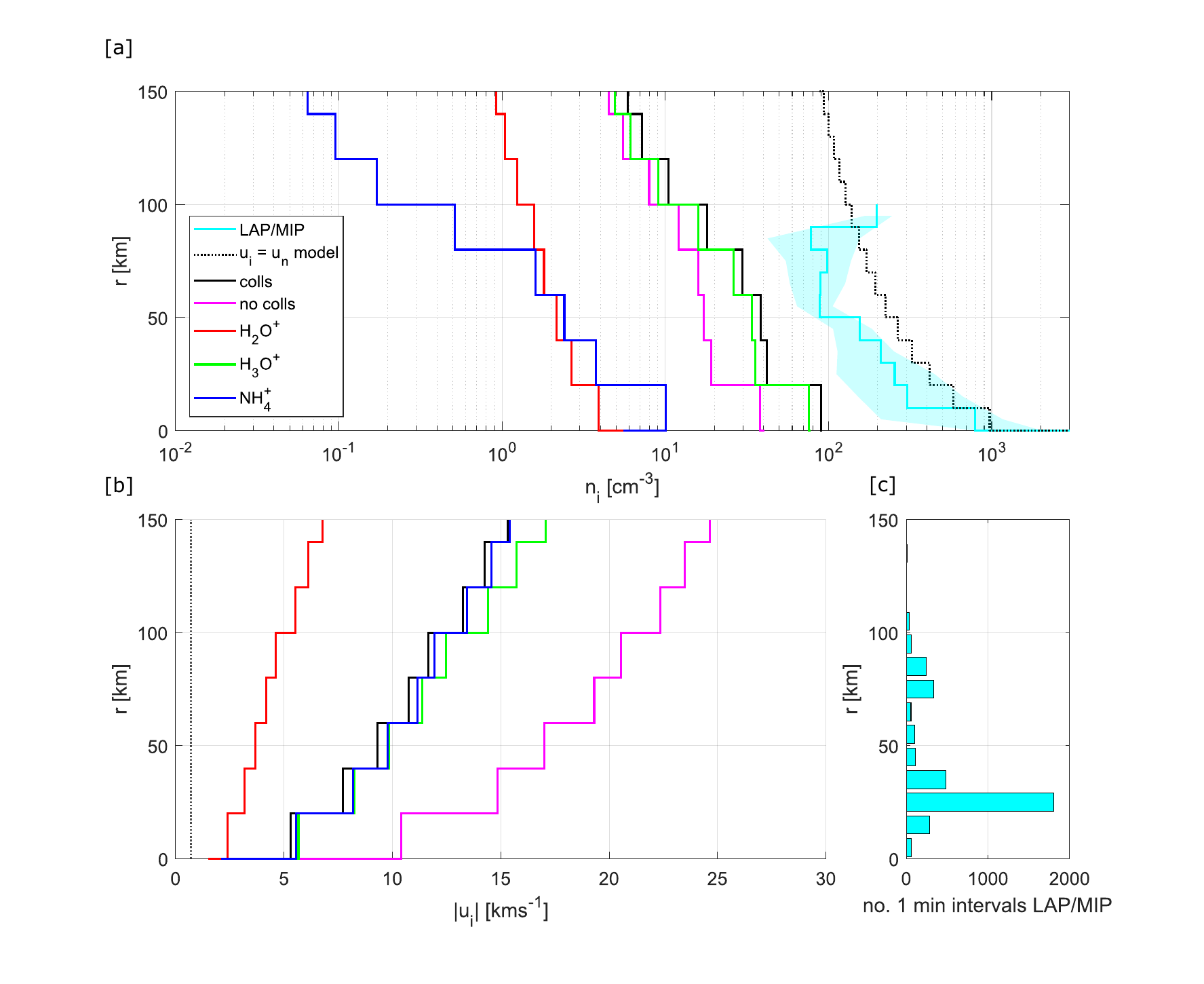}
    \caption{[a] Ion density averaged over $20 \; \mathrm{km}$ cometocentric distance bins for the test-particle model in the terminator plane: Total with collisions ({black}), total without collisions ({pink}), $\mathrm{H_2O^+}$ ({red}), $\mathrm{H_3O^+}$ ({green}) and $\mathrm{NH_4^+}$ ({blue}). The field-free, chemistry-free modelled density using Eq.~\ref{eq: FFCF2} is shown by the {black dashed line}. Cyan line shows the median in each radial bin from MIP/LAP for $19.85 < \log_{10}(Q \nu^{\mathrm{tot}}) < 20.15$. {Shaded region} shows the 25th-75th percentile range of the data in each bin. [b] same as [a] but for the ion bulk velocity from the test-particle model only. [c] Number of 1-minute intervals in each $10~\mathrm{km}$ bin for the MIP/LAP dataset. }
    \label{fig: test pl data comparison}
\end{figure*}

For comparison between the MIP/LAP data and the ion test-particle model run, simulated ions are used from the terminator (Y-Z) plane only. This is where Rosetta made the majority of its orbits around the comet, so is most applicable for data comparison. These simulated data are then averaged over $20 \; \mathrm{km}$ radial bins and compared to the MIP/LAP data as presented in Figure~\ref{fig: ne dens}, for $19.85 < \log_{10}(Q \nu^{\mathrm{tot}}) < 20.15$. The result is shown in Figure~\ref{fig: test pl data comparison}. Even with the collisions implemented, the resulting ion densities are consistently lower than the plasma density measured by the spacecraft, by 5-10 times. The measured densities lie between the $u_i = u_n$ model (Eq.~\ref{eq: FFCF2}) and those calculated by the test-particle model, with good agreement between the data and the Eq.~\ref{eq: FFCF2} particularly below 30 km. The reasons for the too low densities obtained by the test-particle model are discussed in Section~\ref{sec: discussion}, but it is clear that the cometary ions are transported too quickly in the hybrid model.

The ion bulk velocities for each of the species and the total from the test-particle model are shown in Figure~\ref{fig: test pl data comparison}b. It shows that the ions in the model are quickly accelerated to values significantly above the neutral speed, even in the first 20~km from the surface. The inclusion of collisions in the model decreases the bulk ion speed from $20~\mathrm{km~s^{-1}}$ to $11~\mathrm{km~s^{-1}}$ at $100~\mathrm{km}$.  Despite the deceleration due to collisions, these speed values remain high, leading to underestimation of the plasma density compared with the observations. The reason for the high ion speeds is discussed in Section~\ref{sec: chap 5 e temp}.

\section{Discussion}
\label{sec: discussion}

\subsection{Impact of 1D fluid vs 3D kinetic collisions}
\label{sec: chap 5 3D colls discussion}

In Section~\ref{sec: new colls dm cavity}, the 3D test-particle model was applied to the radial electric field derived for the diamagnetic cavity crossings on 21st November 2015 in \citet{Lewis2024}. It was shown that the energy-dependent 3D cross sections make the ion environment more collisional at this outgassing, driving more proton transfer and increasing the prevalence of $\mathrm{NH_4^+}$. However, the total $n_iu_i$ at a given cometocentric distance is preserved such that mean bulk ion velocity and electric field strengths required to reproduce the plasma density observed at Rosetta remain the same. 

For the radial electric field case, the impact of the exothermic energy release during protonation was shown to be small but measurable, altering the total ion bulk speed by up to $8\%$. This is likely because the electric field and momentum transfer collisions quickly `fix' the velocity back to being directed radially outwards from the nucleus. However, the exothermic energy release requires further study as it may have important implications for the ion dynamics at lower outgassing, when the electric field is weaker \citep{Stephenson2024}, though collisions are also less likely in this regime.

\subsection{Application to intermediate outgassing}

Applying the test-particle model to the electric and magnetic fields from the hybrid simulation AMITIS shows that the ion dynamics are different for individual ion species due to their different chemical pathways.

$\mathrm{H_2O^+}$ is close to photochemical equilibrium, even at $2.5-3$~au, so is present in low amounts and not concentrated in any particular location (see Figure~\ref{fig: amitis density}c). Conversely, $\mathrm{H_3O^+}$ and $\mathrm{NH_4^+}$ are more strongly affected by the electric fields, accumulating in greater numbers both near the nucleus and in the ion pile up region inside the magnetic pile up region (see Figure~\ref{fig: amitis density}d-e). Collisions of all types are most effective in this ion pile up region, which extends into the comet tail. The chemistry is less complex in the +Z direction, that is, the direction of the motional electric field where transport is enhanced (see Figure~\ref{fig: amitis fields}). 

 It was previously assumed that the radial flow assumption was only valid in the diamagnetic cavity, when the only electric field term present was the ambipolar electric field. Figure~\ref{fig: amitis quivers} shows that within the first $100\;\mathrm{km}$ of the nucleus, radial motion dominates the flow in the Y-Z plane assessed by Rosetta (for conditions with no diamagnetic cavity). However, comparison with the MIP/LAP data suggests that the ions in the simulation may be accelerated too strongly, likely due to limitations of the hybrid model and the assumptions made about the electron temperature. This is discussed further in the next Section, \ref{sec: chap 5 data comparison discussion}.

 Beyond $100\;\mathrm{km}$, the flow is dominated upstream of the comet by the +Z component, accelerated by the motional electric field of the solar wind. On the anti-sunward side, the flow is tailward (-X), with the velocity direction in the -Z region corresponding to an electric field enhancement. This behaviour at large cometocentric distances is reported in detail by \citet{Moeslinger2024}, so in this study, we focus mainly on the assessment of the cometary ion population at the spacecraft-comet distances covered by Rosetta at $2.5-3$~au. 

\subsection{Comparison of the ion test-particle model and Rosetta data}
\label{sec: chap 5 data comparison discussion}

In Section~\ref{sec: chap 5 data comparison}, the ion test-particle model was used to assess the impact of the electric and magnetic fields from the AMITIS hybrid simulation on the ion composition and density, and compared it to the equivalent plasma density observations from the combined MIP/LAP dataset. The input conditions (Table \ref{tab: hybrid comparisons}) used were representative of those witnessed by Rosetta at $2.5-3$ au post-perihelion. 

The key finding of this comparison is that the plasma density data at this outgassing ($\sim 1.5 \times 10^{26}~\;\mathrm{s^{-1}}$) are underestimated by the test-particle model, even with collisions slowing the acceleration by the electric fields. In fact, the bulk of the data lie somewhere between the profile calculated by the 3D model and the one calculated by the field-free, chemistry-free model (which assumes $u_i = u_i = 0.7 \; \mathrm{km \; s^{-1}}$, Eq.~\ref{eq: FFCF2}, see Figure~\ref{fig: test pl data comparison}). To understand the underestimation of the plasma density by the model, we need to reflect on the assumptions made in both the test-particle model and the hybrid electric and magnetic fields that drive it. 

\subsubsection{Electron-impact ionisation}
\label{sec: chap 5 e-i impact ionisation}

In the hybrid simulations, photoionisation and charge exchange with the solar wind are considered. Electron-impact ionisation is not modelled, only treated by boosting the photoionisation, because its proper assessment requires the kinetic treatment of electrons. 

In an optically thin coma the photoionisation frequency is broadly constant with radial distance but this may not be the case for the electron impact. \citet{Stephenson2023a} modelled the solar wind and cometary electrons using the collisional test-particle model. They showed that the electron-impact ionisation frequency, primarily due to accelerated solar wind electrons and their secondaries, was enhanced at an outgassing of $10^{26}~\mathrm{s^{-1}}$ in the first 100~km from the nucleus. This suggests that simple `boosting' of the photoionisation frequency to account for electron impact (informed by the data from \citet{Stephenson2023a}) is unlikely to capture the full radial profile of the ion production rate properly. 

However, multi-instrument analysis of far-ultraviolet emission using in-situ electron impact frequencies for dissociative excitation (assumed to be independent of cometocentric distance) has shown to produce simulated brightnesses consistent with observations (\citet{Stephenson2021}, \citet{Galand2020}). This suggests that a spatially-uniform electron-impact ionisation frequency, for uniform neutral outgassing, as assumed here, is likely to be a reasonable assumption. 

\subsubsection{Electron temperature}
\label{sec: chap 5 e temp}

The collisionality of electrons observed near the nucleus raises further questions about the validity of the hybrid simulation and its fluid treatment of the electrons. The electrons are assumed to be adiabatic, such that
 \begin{equation}
     p_e =  n_{\text{e}} k_B T_{\text{e}}= \alpha n_e^{\gamma}
     \label{eq: adiabatic elecs}
 \end{equation}
 and 
 \begin{equation}
     T_e = T_{e,SW} \left( \frac{n_e}{n_{e,sw}} \right )^{\gamma - 1}
     \label{eq: adiabatic elec temp}
 \end{equation}
 where $p_e$ is the total thermal pressure of the electrons, and $\alpha  = k_B n_{\text{e},sw}^{1-\gamma} T_{\text{e},sw}$ where $n_{\text{e},sw}$ and $T_{\text{e},SW}$ are the number density and temperature of the electrons in the solar wind, respectively (fixed in the inputs of the simulations). $\gamma$ is the adiabatic index, assumed 5/3 (three degrees of freedom for monatomic gas). In the solar wind, or at low enough neutral density (i.e. sufficiently far from the nucleus), this treatment of the electrons as one population with a Maxwellian distribution may be reasonable, but is clearly an oversimplification in the inner coma. In this region, the cometary electrons have distinct populations according to the collisions and field acceleration they undergo, and, beside a warm $e^-$ population ($\sim 5-10~$eV), a significant population of cold electrons was observed throughout the Rosetta escort \citep{Gilet2020}.

\citet{Stephenson2023a} showed using their 3D electron test-particle model, that the electron-neutral collisions produce cold electrons and a `flattened' potential well in the collisional region surrounding the nucleus. Neglect of this collisional cooling is a serious limitation of the hybrid simulations that are based on the adiabatic temperature assumption. Such an assumption leads to a very high $\vec{E}_{amb}$, hence high ion speeds and low ion densities. For the model runs in Section~\ref{sec: chap 5 data comparison}, a lower ambipolar electric field strength near the surface would result in less acceleration of the cometary ions. That is, slower bulk velocity and higher densities - perhaps closer to the measured plasma densities from MIP/LAP. A similar observation was made in \citet{Koenders2015}, who solved the electron pressure equation in place of the adiabatic assumption, noting by comparison to \citet{Koenders2013} (who used an adiabatic pressure) that the latter caused an additional force on the ions in the inner coma.  

The collisionality of the electrons may therefore be key to understand why the radial $u_i = u_n$ model works so well at low outgassing. However, the impact of the electron-neutral collisions at higher outgassing cannot be determined without a fully kinetic and collisional 3D model of electrons and self-consistent fields.

\section{Conclusions}

In Section~\ref{sec: new colls dm cavity}, we showed that for a radial electric field (and no magnetic field, as inside the diamagnetic cavity) the 3D collisional cross sections lead to up to $20\%$ more collisions than the 1D kinetic rates used in \citet{Lewis2024}. However the same bulk velocity, hence the same ambipolar field, is required to explain the electron density measured by Rosetta. 

By assessing the effect of its inclusion in the ion test-particle model, it was demonstrated that the exothermic energy released in protonation reactions only plays a minor role in the ion dynamics near perihelion. It is possible, however, that this energy release may have more consequences when the electric fields applied are weaker.   

In Section~\ref{sec: application test pl}, the ion test-particle model was applied to input conditions representative of comet 67P at 2.5-3~au, and driven by hybrid electric and magnetic fields produced by the Amitis simulation (see Section~\ref{section: hybrid fields}). For this lower outgassing ($Q=5.4\times 10^{26}~\mathrm{s^{-1}}$), when no diamagnetic cavity has formed, cometary ions still move predominantly radially, up to $100~\mathrm{km}$ above the surface. $\mathrm{H_2O^+}$ is close to photochemical equilibrium, while $\mathrm{H_3O^+}$ and $\mathrm{NH_4^+}$ are strongly affected by the strong ambipolar electric and magnetic fields generated by the hybrid simulation.

Finally, in Section~\ref{sec: chap 5 data comparison} we compare the total plasma density from the ion test-particle model with the in situ data from Rosetta. Even with collisions, the total plasma density derived by the test-particle model is low compared to the MIP/LAP data at the same outgassing. This seems to result from the assumption of adiabatic electrons driving strong ambipolar electric fields in the hybrid simulations, and therefore overestimating the loss of ions through transport. This highlights that kinetic, collisional modelling of the cometary electrons are necessary to understand the ion dynamics and chemistry fully. 

\section*{Acknowledgements}

We would like to acknowledge the invaluable work of the RPC team, the whole ESA Rosetta team, and the ESA Planetary Science Archive team.  Work at Imperial College London was supported by the Science and Technology Facilities Council (STFC) of the UK under studentship ST/W507519/1 and grant ST/W001071/1, and by the UK Space Agency (UKSA) under grant ST/X002349/1. 

\section*{Data Availability}


The data used in this article are available on the Planetary Science Archive at \mbox{\url{https://psa.esa.int}}. The AMITIS simulation results used in this work are available at \citet{Moeslinger2024a}.



\bibliographystyle{mnras}
\bibliography{references} 




\appendix

\section{Validation of the test-particle model}

\subsection{Radial electric fields with collisions}
\label{sec: radial fields validation appendix}
\begin{figure}
    \includegraphics[width=\linewidth]{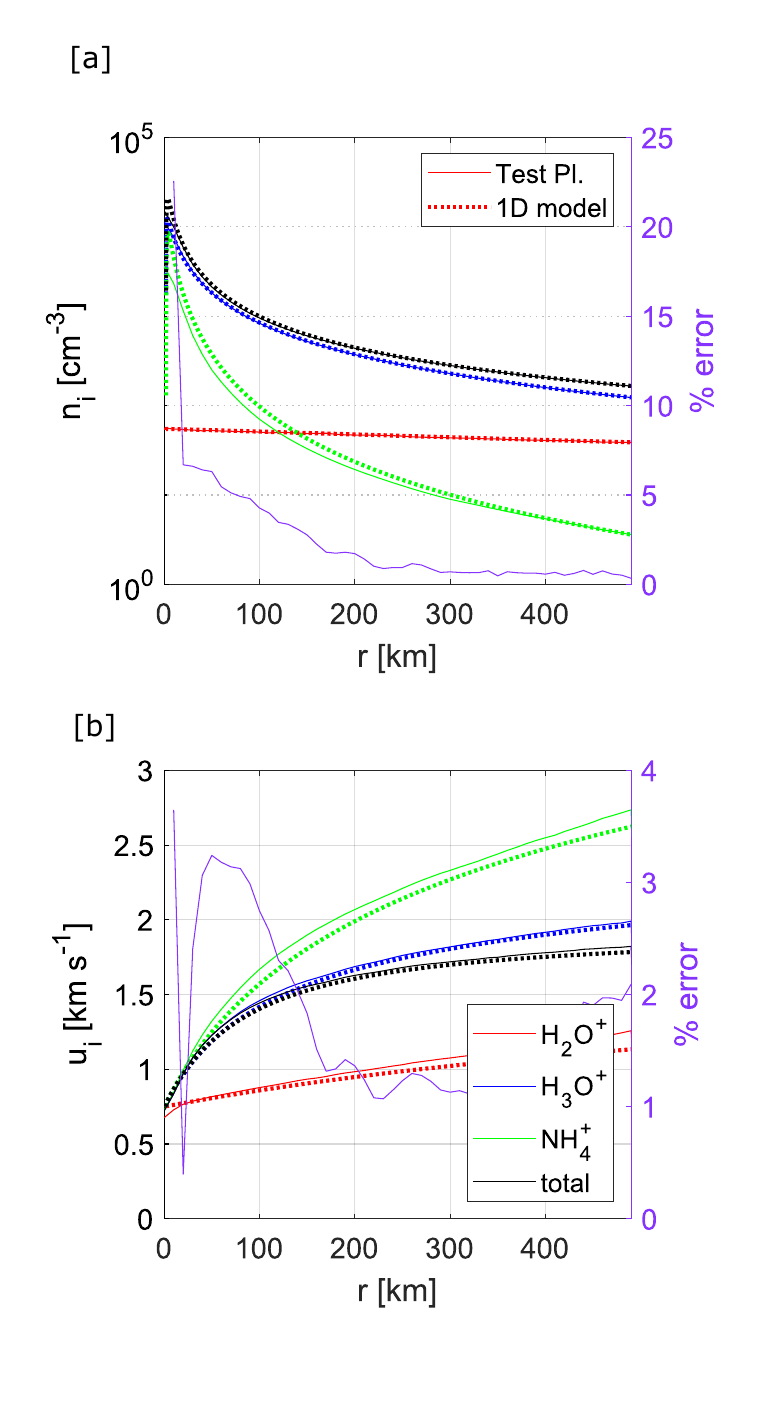}
    \caption{[a] $\mathrm{H_2O^+}$ (red),  $\mathrm{H_3O^+}$ (blue), $\mathrm{NH_4^+}$ (green) and total (black) density profiles from the 1D Ion Acceleration Model described in \citet{Lewis2024} (dotted lines) compared with those from the test-particle model (solid lines). [b] same as [a] but for the bulk ion velocity. Input parameters are given in Table \ref{tab: validation parameters}}
    \label{fig: validation no hybrid}
\end{figure}

Here the collisional aspect of the ion test-particle model is validated, without the hybrid electric and magnetic field input. Figure \ref{fig: validation no hybrid} shows the comparison of the ion test-particle model (Section~\ref{sec: modelling}) with the 1D ion acceleration model presented in \citet{Lewis2024}. Models are both run using the same electric field profile (radial and proportional to $1/r$) and using the input parameters given in Table~\ref{tab: validation parameters}. 

\subsection{Collisionless case}
\label{sec: collisionless validation appendix}
\begin{figure*}
    \includegraphics[width=0.9\linewidth]{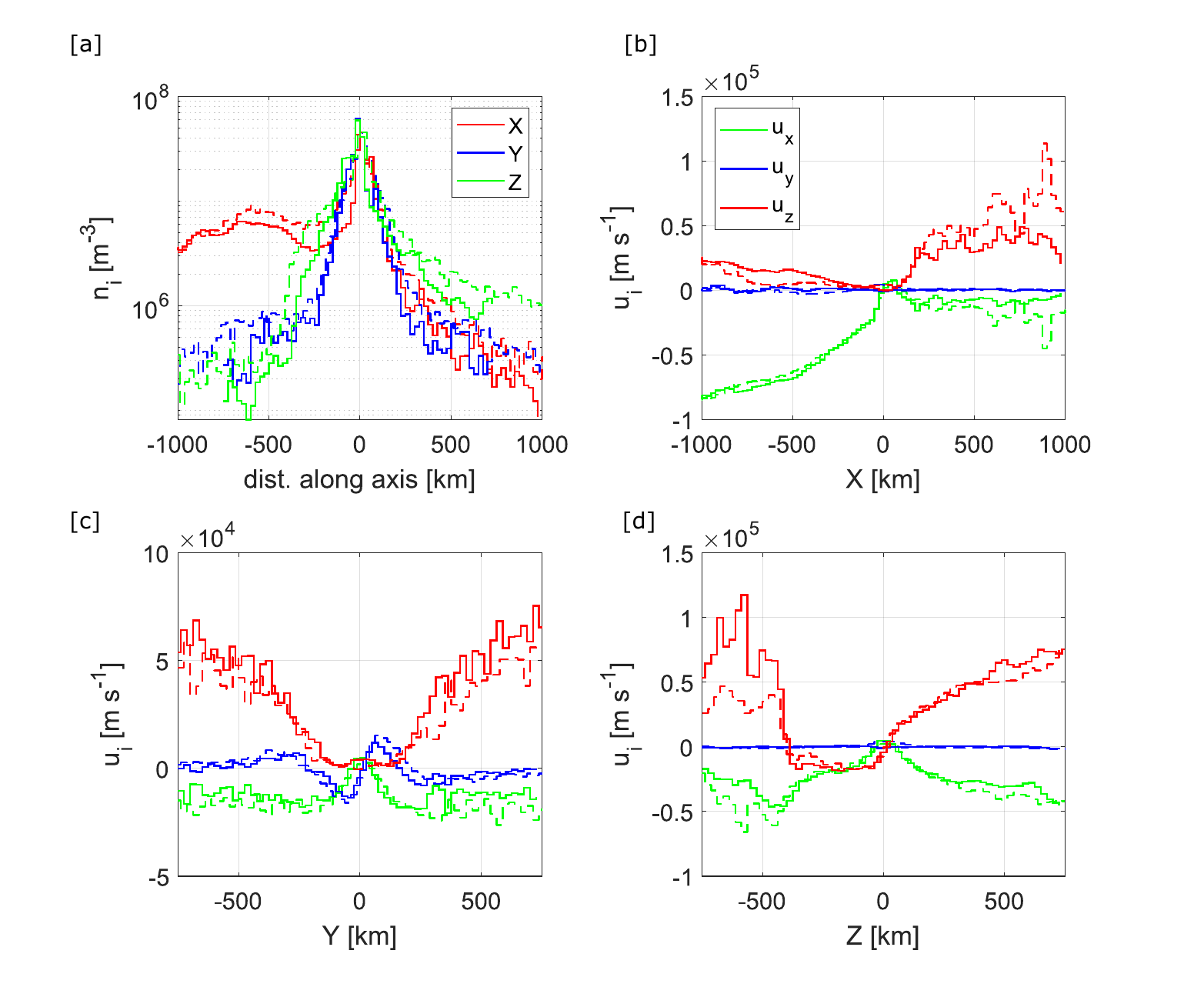}
    \caption{Comparison of the test-particle model described in Section~\ref{sec: test pl description} with collisions removed (solid lines), and the AMITIS collisionless hybrid simulation (dashed lines). [a] Total ion density along the X, Y and Z axis. [b-d], X (green), Y (blue) and Z (red) components of the bulk ion velocity along each axis. The sun is in the +X direction, with the interplanetary magnetic field in the +Y direction.}
    \label{fig: validation no colls}
\end{figure*}

In Figure~\ref{fig: validation no colls}, The test-particle model is now validated in the collisionless case against the output cometary ion density from AMITIS \citep{Moeslinger2024}. The input parameters are given in Table~\ref{tab: hybrid comparisons}.

\section{Ion-neutral collision cross sections}
\label{sec: test particle cross secs appendix}

\begin{table*}
\begin{tabular}{l|l|l|l|l}
\textbf{Reaction type}              & \textbf{Equation} & \textbf{Cross section } $[10^{-16}~\mathrm{cm^2}]$ & \textbf{Range [eV]} & \textbf{Reference} \\[5pt] \hline
Proton transfer           &    $\mathrm{H_2O^+} + \mathrm{H_2O} \rightarrow \mathrm{H_3O^+} + \mathrm{OH}$      &    $\sigma(E_{\mathrm{rel}}) = 38 E_{\mathrm{rel}}^{-0.88}- 0.39\exp{\left[-0.5 (\frac{E_{\mathrm{rel}} - 57}{12})^2\right]}$           &    0.1 - 100          &   \cite{Fleshman2012}        \\ [5pt]

Proton transfer           &    $\mathrm{H_2O^+} + \mathrm{NH_3}\rightarrow \mathrm{NH_4^+} + \mathrm{OH}$      &    $\sigma(E_{\mathrm{rel}}) = 38 E_{\mathrm{rel}}^{-0.88}- 0.39\exp{\left[-0.5 (\frac{E_{\mathrm{rel}} - 57}{12})^2\right]}$           &    0.1 - 100          &   \cite{Fleshman2012}        \\[5pt]

Proton transfer            &    $\mathrm{H_3O^+} + \mathrm{NH_3}\rightarrow \mathrm{NH_4^+} + \mathrm{H_2O}$      &    $\sigma(E_{\mathrm{rel}}) = 38 E_{\mathrm{rel}}^{-0.88}- 0.39\exp{\left[-0.5 (\frac{E_{\mathrm{rel}} - 57}{12})^2\right]}$           &    0.1 - 100          &   \cite{Fleshman2012}        \\ [5pt]

Momentum transfer     &    $\mathrm{H_2O^+} + \mathrm{H_2O} \rightarrow \mathrm{H_2O^+} + \mathrm{H_2O}$      &   $\sigma(E_{\mathrm{rel}}) = 24E^{-0.5}$           &    Not given          &   \cite{Vigren2017}        \\ [5pt]

Momentum transfer           &    $\mathrm{H_3O^+} + \mathrm{H_2O}\rightarrow \mathrm{H_3O^+} + \mathrm{H_2O}$      &   $\sigma(E_{\mathrm{rel}}) = 24E_{\mathrm{rel}}^{-0.5}$           &    Not given         &   \cite{Vigren2017}        \\[5pt]

Momentum transfer           &    $\mathrm{NH_4^+} + \mathrm{H_2O}\rightarrow \mathrm{NH_4^+} + \mathrm{H_2O}$      &   $\sigma(E_{\mathrm{rel}}) = 24E_{\mathrm{rel}}^{-0.5}$           &    Not given          &   \cite{Vigren2017}        \\ [5pt]

Electron transfer           &    $\mathrm{H_2O^+} + \mathrm{H_2O}\rightarrow \mathrm{H_2O^+} + \mathrm{H_2O}$      &   $\sigma(E_{\mathrm{rel}}) = 38E_{\mathrm{rel}}^{-0.5}$     &    0.1 - 100          &   \cite{Fleshman2012}        \\ \hline

\end{tabular}
\label{tab: test pl cross secs}
\caption{Ion-neutral collision cross sections used in the test-particle model}
\end{table*}


\bsp	
\label{lastpage}
\end{document}